\documentclass[preprint,
3p,
times loads txfonts
]{elsarticle}

\usepackage{graphicx}% Include figure files
\usepackage{dcolumn}% Align table columns on decimal point
\usepackage{bm}% bold math
\usepackage{subcaption}

\usepackage[utf8]{inputenc}
\usepackage[T1]{fontenc}
\usepackage{mathptmx}
\usepackage{etoolbox}
\usepackage{dsfont}
\usepackage{xcolor}
\usepackage{bigints}
\usepackage{relsize}
\usepackage{exscale}
\usepackage{amsmath}
\usepackage{amssymb}
\usepackage{wrapfig}
\usepackage{hyperref}
\usepackage{comment}
\usepackage[export]{adjustbox}
\usepackage{orcidlink}

\newcommand{\B}{\ensuremath{\bm{B}}}
\newcommand{\lb}{\ensuremath{\bm{l}}}

\newcommand{\E}{\ensuremath{\bm{E}}}
\newcommand{\n}{\ensuremath{\bm{\hat{n}}}}
\newcommand{\rr}{\ensuremath{\bm{r}}}
\newcommand{\rp}{\ensuremath{\bm{r'}}}
\newcommand{\G}{\ensuremath{\mathcal{G}(\rr,\rp)}}
\newcommand{\Gint}{\ensuremath{G(\rr,\rp)}}

\newcommand{\half}{\ensuremath{\frac{1}{2}}}
\newcommand{\threehalf}{\ensuremath{\frac{3}{2}}}

\newcommand{\x}{\ensuremath{\bm{x}}}

\newcommand{\xtor}{\ensuremath{\bm{x}_\mathrm{tor}}}
\newcommand{\er}{\ensuremath{\bm{\hat{e}}_R}}
\newcommand{\ez}{\ensuremath{\bm{\hat{e}}_Z}}
\newcommand{\ephi}{\ensuremath{\bm{\hat{e}}_\phi}}
\newcommand{\ex}{\ensuremath{\bm{\hat{e}}_x}}
\newcommand{\ey}{\ensuremath{\bm{\hat{e}}_y}}

\newcommand{\xipph}{\ensuremath{\xi_{2,i} + \frac{\Delta\xi_2}{2}}}
\newcommand{\xitph}{\ensuremath{\xi_{3,i} + \frac{\Delta\xi_3}{2}}}
\newcommand{\xipmh}{\ensuremath{\xi_{2,i} - \frac{\Delta\xi_2}{2}}}
\newcommand{\xitmh}{\ensuremath{\xi_{3,i} - \frac{\Delta\xi_3}{2}}}
\newcommand{\lftp}{\ensuremath{{\xi_{2,i}}^L}}
\newcommand{\rgtp}{\ensuremath{{\xi_{2,i}}^R}}
\newcommand{\lftt}{\ensuremath{{\xi_{3,i}}^L}}
\newcommand{\rgtt}{\ensuremath{{\xi_{3,i}}^R}}
\newcommand{\intxip}{\ensuremath{\bigintssss_{\mathlarger{\lftp}}^{\mathlarger{\rgtp}}}}
\newcommand{\intxit}{\ensuremath{\bigintssss_{\mathlarger{\lftt}}^{\mathlarger{\rgtt}}}}
\newcommand{\phitor}{\ensuremath{{\phi_\mathrm{tor}}}}

\makeatletter
\def\@email#1#2{%
 \endgroup
 \patchcmd{\titleblock@produce} using 
  {\frontmatter@RRAPformat}
  {\frontmatter@RRAPformat{\produce@RRAP{*#1\href{mailto:#2}{#2}}}\frontmatter@RRAPformat}
  {}{}
}%
\makeatother
\begin{document}

%\preprint{AIP/123-QED}

\begin{frontmatter}

\title{Implementation and Verification of Toroidal Resistive Wall Boundary \\ Conditions in the PIXIE3D MHD code using a Boundary Integral Method}

\author[1]{S. W. Jones\corref{cor1}\orcidlink{0000-0003-3970-1843}}\ead{swj@lanl.gov}
\author[1]{L. Chac{\'o}n\orcidlink{0000-0002-4566-8763}}
\author[1]{J. Hamilton\orcidlink{0000-0003-1289-8180}}
\author[2]{D. C. Barnes}
\author[1]{A. Yag\"ue-L\'opez\orcidlink{0000-0002-7294-9288}}

\address[1]{Los Alamos National Laboratory, Los Alamos, NM, 87545, USA}
\address[2]{Coronado Consulting, Lamy, NM 87540}
\cortext[cor1]{Corresponding Author}

\begin{abstract}
We present the complete formulation of resistive wall boundary conditions in axisymmetric toroidal geometries as implemented in the PIXIE3D extended magnetohydrodynamics (MHD) code, along with a complete suite of analytical verification examples that demonstrate correctness in the implementation. The formulation centers around a thin wall approximation and a Boundary Integral Method to solve for the magnetic scalar potential in the immediately surrounding vacuum. This requires specialized quadrature rules derived from existing literature to handle the numerical integration of singular and hypersingular integrands (the Green's function of Laplace's equation and its derivatives), for which we provide the nodes and weights.
Further, we describe an extension to the formalism to include the effect of a perfectly conducting second, outer wall exterior to the resistive (plasma-facing) wall and separated by vacuum, and exterior to the computational plasma mesh proper. Lastly, we describe an extension to include the effect of current-carrying coils also defined exterior to the plasma mesh in the resistive wall boundary condition treatment.
For most aspects of the method, we present self-contained verification examples using analytic solutions in axisymmetric toroidal geometries (with both 2D and 3D plasmas) and show it to be accurate to second order.
We demonstrate the algorithm with a vertical displacement event (VDE) using the ITER tokamak geometry.
\end{abstract}

\begin{keyword}
Resistive wall boundary \sep Magnetohydrodynamics \sep Magnetic fusion \sep boundary integral methods \sep numerical verification
\vspace{.5\baselineskip}
%\MSC 65M12 \sep 65M60 
\end{keyword}

\end{frontmatter}
%\maketitle

\section{\label{sec:introduction}Introduction}

Tokamak disruptions are events in which confinement of the plasma and plasma current is rapidly lost, typically on the order of one to ten milliseconds. Disruptions are triggered by the growth of MHD instability modes such as vertical displacement events (VDEs), kink, tearing or resistive wall (RWM) modes. Strong cooling from bremsstrahlung or impurity recombination radiation can also result in a disruption.

Both RWMs and VDEs involve finite resistivity of the plasma wall. This so-called ``resistive wall'' (RW) allows magnetic field to diffuse through, potentially destabilizing the confined plasma. VDEs occur when magnetohydrostatic equilibrium of the plasma is lost, leading the plasma column to drift vertically in the device \cite{gruber1993vertical}. RWMs are 3D magnetohydrodynamic (MHD) modes destabilized by the magnetic field  penetration of the RW allowed by finite wall resistivity \cite{igochine2012physics}.

The consequences of a disruption, other than loss of confinement and energy, are strong electromagnetic forces on the containment vessel, runaway electrons and strong wall heating, all of which are very undesirable and can lead to irreversible damage to the device. Therefore, understanding and mitigating disruptions is crucial for designing performant tokamak reactors.
The RW is a key physics component needed for such understanding. The RW boundary treatment in MHD is not new, and most modern MHD simulation tools for magnetic fusion devices feature a suitable RW treatment to connect effectively with experiments~ (see e.g.~\cite{marx2017a,krebs2020a}). However, implementation and verification of an MHD boundary RW treatment is generally quite inaccessible for the novice practitioner, due to the complex geometries involved and the need to couple with external libraries to read suitable MHD equilibria in such geometries (see e.g. the recent verification exercise in \cite{krebs2020a}). This study has a double aim: 1) to document the specific implementation details of a thin-wall RW boundary module in toroidal geometry using a Boundary Integral Method (BIM) in the PIXIE3D MHD code \cite{chacon2004a,chacon2008a,chacon2025a,hamilton2025a}, and 2) to document several self-contained analytical RW verification tests in accessible but realistic geometries that can be readily implemented without need of external setup information.

The formulation of RW models as boundary conditions for MHD simulations is fairly mature, with abundant literature. A common (and efficient) approximation for the RW treatment is the so-called ``thin-wall'' approximation, which treats the RW as a thin resistive layer of finite (but small) width, and ignores the current distribution across the layer. The formulation of thin-wall RW models is mature and well known.
The first report of a RW boundary treatment with MHD is in \cite{schnack1990computational} for a Reversed Field Pinch in cylindridal geometry. More complicated geometries were considered later by \cite{chance1997a}, who used the BIM \cite{hall1994a} for vacuum solutions to Laplace's equation in axially symmetric toroidal geometry for the vacuum magnetic scalar potential surrounding a plasma discharge, including in the presence of external conductors. The authors noted that the external conductor should be sufficiently far from the plasma surface or any other conducting surface in order for the method to be accurate. That approach was later extended in \cite{chance2007a}, which presented a method for accurately integrating the Green's function for the Laplacian operator in axially symmetric geometry, which is robust for large toroidal mode numbers $n$. The inclusion of RW effects in the determination of MHD equilibria was carefully considered in \cite{pustovitov2023a}.

Various production codes have documented detailed implementation of RW modules. In \cite{marx2017a}, simulations of tokamak plasma instabilities using the XTOR-2F code were presented, using a free-boundary approach with or without a thin resistive wall. They use the GRIN software package \cite{pletzer2011a,grin} to generate vacuum response matrices to compute the magnetic potential and therefore construct the boundary condition. GRIN constructs the vacuum response matrices using regularization and operator transformation techniques to transform a Poisson-type equation into a Helmholtz-type equation. In their paper \cite{pletzer2011a}, they perform verification of the vacuum response for a cylindrical boundary with a single Fourier mode and also for a Grad-Shafranov equilibrium. GRIN is also used by the M3D MHD code \cite{park1999a,strauss2023a}. Further verification of the XTOR-2F code was detailed in \cite{marx2017a}, which presented simulations for an axially symmetric mode instability and an external kink mode and associated resistive wall instability. The authors compared against analytic linear growth rates for large aspect ratio tokamaks. 
NIMROD \cite{sovinec2004a} is another code that uses GRIN to compute vacuum response matrices for a general resistive wall boundary condition \cite{becerra2016a}. They have conducted verification for a periodic cylinder and shown the correct resistive wall mode growth rates compared to analytic growth rates for a large aspect ratio torus. The implementation was found to converge consistently at second order.
The study in \cite{spinicci2023a} presented a careful MHD+RW cross-code verification between the Specyl and PIXIE3D codes and with analytical theory for helical and cylindrical geometries; both codes employ a thin-wall approximation. The PIXIE3D implementation in the reference is identical to the one employed in this study.
A different approach is taken by \cite{ferraro2016a} in the code M3D-C1, whereby the resistive wall and surrounding vacuum is included in the computational mesh. The authors apply this to simulate external kink resistive wall modes, and verify their implementation against analytic growth rates for resistive wall modes in cylindrical geometry \cite{liu2008a}.
The JOREK code \cite{czarny2008a} is coupled to a modified version of the STARWALL code \cite{merkel2015a}, which computes the vacuum response matrices in the presence of a three-dimensional conducting external structure with holes. Careful cross-code verification between the NIMROD, JOREK, and M3D-C1 codes for an NSTX equilibrium was documented in \cite{krebs2020a}. 

In this study, we document the implementation of a 3D toroidal RW boundary module in the PIXIE3D MHD code. PIXIE3D~\cite{chacon2008a} is a modern MHD simulation code for MFE devices, based on fully implicit timestepping algorithms and fast multigrid-preconditioned Jacobian-free Newton-Krylov solvers. PIXIE3D already features many of the ingredients necessary to model key experimental features of a tokamak disruption  \cite{hamilton2025a} including nonlinear (Braginskii) transport coefficients, the ability to deal with strong anisotropic transport \cite{chacon2025b}, and Hall effects \cite{chacon2025a}. PIXIE3D has been carefully verified against theory and other MHD codes in fusion-relevant contexts using ideal walls \cite{bonfiglio2010nonlinear}. A RW boundary module based on a thin-wall approximation has been recently implemented in PIXIE3D, also carefully verified for helical and cylindrical geometries in~\cite{spinicci2023a}.

In what follows, we document the generalization of the vacuum-solution in PIXIE3D to axisymmetric toroidal geometries (but still allowing for 3D plasma evolution). Our approach avoids gridding up the vacuum region by employing a BIM to find the magnetic potential solution directly at the vacuum side of the plasma wall. BIM is in principle a very efficient method because only requires meshing up the surface of interest, but is highly singular and requires extremely careful numerics. The use of BIM methods for RW boundary conditions in MHD codes is by itself not new (see e.g. \cite{chance1997a,chance2007a,marx2017a,becerra2016a}). The main contribution of this study, beyond documenting the toroidal RW treatment in PIXIE3D, is a complete suite of self-contained analytical verification tests for the vacuum magnetic field solver, which are approachable for any practitioner interested in implementing similar physics with low geometric complexity and without needing the availability of other fully featured MHD codes. Such as suite allows for a careful characterization of numerical convergence of the algorithm for realistic toroidal geometries, and to the best of our knowledge is missing in the literature. We also carefully document every component of the algorithm, including high-order quadrature weights for hypersingular integrals, for the benefit of future practitioners.

The rest of this paper is organized as follows. We first introduce the BIM for obtaining the scalar magnetic potential in the vacuum surrounding the wall in a general coordinate system (Section~\ref{sec:theory-foundation}). 
Our numerical discretization of the boundary integral equation in PIXIE3D's generalized curvilinear coordinate system is presented in Section~\ref{sec:numerical-method}. Sections~\ref{sec:external_wall}~\&~\ref{sec:external_coils} describe the extensions required to include either a second external conducting wall or a set of external control coils. The method is verified in Section~\ref{sec:verification}, and we present a VDE as an application example in Section~\ref{sec:VDE-ITER}. %~\&~\ref{sec:VDE-NSTX}. 
Finally, we conclude in Section~\ref{sec:conclusions}.

\section{\label{sec:theory-foundation}Theoretical Foundations and Boundary Integral Method}

\subsection{Resistive wall boundary condition in PIXIE3D}

PIXIE3D \citep{chacon2004a,chacon2008a,chacon2025a} is a 3D extended MHD code in generalized curvilinear coordinates. It uses a finite difference discretization of the extended MHD equations in space, which is integrated in time using a first order Jacobian-Free Newton-Krylov method. It is fully parallelized using MPI and PETSc \cite{petsc-efficient}. This work is concerned specifically with the induction equation:
\begin{align}
    \dfrac{\partial\B}{\partial t} + \nabla\times\E = 
    \dfrac{\partial\B}{\partial t}
    + \nabla\times(-\bm{v}\times\B 
    + \dfrac{\eta}{\mu_0}\nabla\times\B) = 0,
\end{align}
for which a boundary condition for $\E$ is clearly required in order to evolve $\B$.
PIXIE3D employs a resistive wall boundary condition for $\E$ using a thin-wall approximation, which is described in \cite{spinicci2023a} and first introduced in \cite{schnack1990computational}. We repeat the key details here for completeness but refer the interested reader to those studies for further details.

The first wall is assumed to be a thin resistive shell of thickness $\delta_w$ at $r=a$, which obeys Ohm's law
\[
\E_w=\eta_w\mathbf{J}_w.
\]
Ohm’s and Ampère’s laws are integrated across the wall, giving the electric field in the wall as
\[
\E_w=\E_0+\frac{a}{\tau_w}
\n\times\delta\B_t,
\]
where $\E_0$ is some externally imposed electric field, $\n$~is the outward-pointing unit normal vector to the boundary (i.e.~pointing into the plasma domain), and $\tau_w$ is the resistive wall time
\[
\tau_w=\frac{\mu_0v_A\tau_A\delta_w}{\eta_w}.
\]
In the limit $\tau_w\ll\tau_A$, the wall behaves as a `transparent' vacuum wall, and in the limit $\tau_w\gg\tau_A$, the wall behaves like a perfect conductor.
The question at hand is how to calculate the jump in the tangential magnetic field
\[\delta\B_t = [\B_t]_-^{+}\equiv\B_t^+ - \B_t^-,\]
by computing the tangential component of the magnetic field $\B_t^+$ from the known magnetic field on the plasma side $\B^-$. One very efficient solution to this problem, proposed in the literature, is to use the boundary integral method (BIM) to compute the vacuum response to the plasma magnetic field. We describe this next.

\subsection{Vacuum response and boundary integral method}
Magnetically confined thermonuclear plasmas in most experimental fusion facilities are surrounded by a vacuum chamber with finite resistivity. In the ITER tokamak, the first-wall panels are made from tungsten (although earlier designs called for beryllium) and the vacuum vessel from stainless steel. Separating the two walls is a shielding blanket region also made from stainless steel but with water cooling and a copper alloy heat sink. Optionally, there are lithium test blanket modules for tritium breeding. Although clearly not vacuum, the blanket has sufficiently higher resistivity than either the plasma, first wall or vessel wall that we make the simplifying assumption that the blanket region is vacuum (infinite resistivity). We also make the assumption that the ambient-pressure air region outside of the first wall of NSTX that contains the poloidal field coils is a vacuum for the same reason. 

In the vacuum region surrounding a thin resistive wall, the current density $\bm{J}=0$ and therefore the magnetic field is irrotational ($\nabla\times\bm{B}=0$) and can be expressed as the gradient of a scalar potential $\bm{B}=\nabla U$. Together with the solenoidal constraint $\nabla\cdot\bm{B}=0$, this gives a Laplacian equation for the magnetic field in the vacuum and a Neumann boundary condition for the scalar potential,
\begin{align*}
    \nabla^2U &= 0, \,\, \mathbf{r} \in \Omega_{vac},\\
    \n\cdot\nabla U& = \n\cdot\B, \,\, \mathbf{r} \in \partial \Omega_{vac}.
\end{align*}
The normal component of the magnetic field must be continuous across the boundary owing to the solenoidal nature of the field. Therefore, $\n\cdot\B$ at the boundary can be computed from the field on the interior of the simulation domain adjacent to the boundary. In order to calculate the tangential components of $\B$~(and, hence, \E) at the boundary, we must solve the Laplace equation for $U$ on the vacuum side of the wall. 

As other practitioners \cite{chance1997a,chance2007a,marx2017a,becerra2016a}, we adopt the boundary integral method to obtain $U$ at the wall. The form of the boundary integral equation for $U$ at the wall (see~\ref{sec:BIM} for a derivation) is
\begin{align}
    \int_B
    U(\rr)\n\cdot\nabla \G dS(\rr)
    -
    \int_B\G\n\cdot\nabla U(\rr)dS(\rr)
    = \dfrac{U(\rp)}{2},
    \label{eq:boundary_integral}
\end{align}
where the free space Green's function is given by
\begin{align}
    \label{eq:freespace-greens}
    \G = -\dfrac{1}{4\pi|\rr-\rp|}.
\end{align}
Specializing to cylindrical $(R,\phi,Z)$ coordinates, we give expressions for the Green's function and its derivatives in \ref{sec:greens}. Using these expressions, the boundary integral equation (\ref{eq:boundary_integral}) becomes
\begin{align}
% cylindrical
    \int
    U(R,\phi,Z)
    \left[
    (\n\cdot\er)
    \dfrac{\partial \G}{\partial R}
    +
    %\right.\nonumber\\
    %\left.
    (\n\cdot\ez)
    \dfrac{\partial \G}{\partial Z}
    \right]
    R dR d\phi dZ  \nonumber\\
    - \int
    J(R)\n\cdot\B(R,\phi,Z)
    \G ~ R dR d\phi dZ
    = \dfrac{U(R',\phi',Z')}{2} &. \label{eq:bimRZ3d}
\end{align}

\section{\label{sec:numerical-method}Numerical Methodology in PIXIE3D}
The PIXIE3D code \cite{chacon2008a} uses a generalized curvilinear mesh with a logically Cartesian coordinate system
$
    \bm{\xi} = (\xi_1,\xi_2,\xi_3), 
$
where $\xi_1$ is the distance from the focal ring and $\xi_2$ and $\xi_3$ are the poloidal and toroidal coordinates, respectively.
We define the normal and tangential bases as $\bm{e}_i=\partial\rr/\partial\xi_i$ and $\bm{e}^i=\nabla\xi_i$, respectively.
The curvilinear mesh is defined such that $\xi_3=\phi_\mathrm{tor}$ and the surface normal is aligned with $\bm{e}^1$ and therefore we write
\begin{align}
    \label{eq:pixie3d-normal}
    \bm{n}\equiv\bm{e}^1 = \nabla\xi_1 = \dfrac{1}{J} \left(\bm{e}_2\times\bm{e}_3\right),
\end{align}
where $\bm{n}$ is the un-normalized normal vector.
Furthermore, since we only consider axisymmetric geometries, the Jacobian and normal vectors are independent of $\xi_3$ and depend only on $\xi_2$ (because $\xi_1$ is a constant at the boundary).
The physical surface area element is thus defined by
\begin{align}
    dS = \left| \bm{e}_2 \times \bm{e}_3 \right|
    d\xi_2d\xi_3
    = J(\xi_2)|\bm{n}(\xi_2)|d\xi_2d\xi_3.
\end{align}
We transform the integrals in (\ref{eq:bimRZ3d}) from cylindrical coordinates to curvilinear coordinates while keeping the expressions for Green's function and its derivatives in cylindrical coordinates. This just results in a change in the integration variables and the surface Jacobian:
\begin{align}
    \int\int
    U(\xi_2,\xi_3)
    J(\xi_2)
    \left[
    (\bm{n}(\xi_2)\cdot\er)
    \dfrac{\partial 
    \mathcal{G}(\xi_2,\xi_3;\xi_{2,j},\xi_{3,j})
    }{\partial R(\xi_2)}
    +
    \right.\nonumber\\
    \left.
    (\bm{n}(\xi_2)\cdot\ez)
    \dfrac{\partial 
    \mathcal{G}(\xi_2,\xi_3;\xi_{2,j},\xi_{3,j})
    }{\partial Z(\xi_2)}
    \right]
    d\xi_2 d\xi_3 \nonumber\\
    - \int\int
    B^n(\xi_2,\xi_3)
    \mathcal{G}(\xi_2,\xi_3;\xi_{2,j},\xi_{3,j})
    d\xi_2 d\xi_3
    - \dfrac{U(\xi_{2,j},\xi_{3,j})}{2} &= 0.
    \label{eq:greens-curvilinear}
\end{align}
Here, the index $j$ assumes a lexicographic ordering of the surface cells discretizing the 2D boundary, and $B^n$ is the contravariant component of the magnetic field normal to the resistive wall surface, which is defined as:
\begin{align}
    \label{eq:bn-definition}
    \dfrac{B^n}{J|\bm{n}|} = \dfrac{\bm{n}}{|\bm{n}|}\cdot\nabla U = \n\cdot\nabla U.
\end{align}
In order to solve (\ref{eq:greens-curvilinear}) numerically, we discretize the integrals in $\xi_2,\xi_3$ as:
\begin{align}
% cylindrical
    \sum_i\intxip\intxit
    U(\xi_2,\xi_3)
    J(\xi_2)
    \left[
    (\bm{n}(\xi_2)\cdot\er)
    \dfrac{\partial 
    \mathcal{G}(\xi_2,\xi_3;\xi_{2,j},\xi_{3,j})
    }{\partial R(\xi_2)}
    +
    \right.\nonumber\\
    \left.
    (\bm{n}(\xi_2)\cdot\ez)
    \dfrac{\partial 
    \mathcal{G}(\xi_2,\xi_3;\xi_{2,j},\xi_{3,j})
    }{\partial Z(\xi_2)}
    \right]
    d\xi_3 d\xi_2 \nonumber\\
    - \sum_i\intxip\intxit
    B^n(\xi_2,\xi_3)
    \mathcal{G}(\xi_2,\xi_3;\xi_{2,j},\xi_{3,j})
    d\xi_3 d\xi_2
    - \dfrac{U(\xi_{2,j},\xi_{3,j})}{2} &= 0,
\end{align}
where the index $i$ assumes the same lexicographic ordering as $j$, and the integral limits are defined as
\begin{align*}
    \lftp = \xipmh, \quad &\rgtp = \xipph, \\
    \lftt = \xitmh, \quad &\rgtt = \xitph.
\end{align*}
Further, we make the second-order approximation of considering $U$, $B^n$ constant per cell, yielding:
\begin{align}
% cylindrical
    \sum_i
    U_i
    \intxip\intxit
    J(\xi_2)
    \left[
    (\bm{n}(\xi_2)\cdot\er)
    \dfrac{\partial 
    \mathcal{G}(\xi_2,\xi_3;\xi_{2,j},\xi_{3,j})
    }{\partial R(\xi_2)}
    +
    \right.\nonumber\\
    \left.
    (\bm{n}(\xi_2)\cdot\ez)
    \dfrac{\partial 
    \mathcal{G}(\xi_2,\xi_3;\xi_{2,j},\xi_{3,j})
    }{\partial Z(\xi_2)}
    \right]
    d\xi_3 d\xi_2 \nonumber\\
    - \sum_i
    B^n_i
    \intxip\intxit
    \mathcal{G}(\xi_2,\xi_3;\xi_{2,j},\xi_{3,j})
    d\xi_3 d\xi_2
    - \dfrac{U_j}{2} &= 0,
\end{align}
where we have used the notation $U_i=U(\xi_{2,i},\xi_{3,i})$ for the pointwise values of $U$ at the face centers.
This forms a linear system
\begin{align*}
    (K_{ji} - \half\delta_{ji})U_i - L_{ji}B^n_i &= 0,
\end{align*}
where Einstein summation convention is assumed, which is solved as
\begin{align}
    U_i &= (K_{ji} - \half\delta_{ji})^{-1}L_{ji}B^n_i.
    \label{eq:linear_system}
\end{align}
in which $(J\bm{n}\cdot\nabla U)=B^n$
is the Neumann boundary condition and $\delta_{ji}$ is the identity matrix.

For clarity, the elements of the matrices $K_{ij}$ and $L_{ij}$ are
\begin{align*}
    K_{ji} &= \intxip\intxit J(\xi_2) \bm{n}(\xi_2)\cdot\nabla
    \mathcal{G}(\xi_2,\xi_3;\xi_{2,j},\xi_{3,j})
    ~d\xi_3 d\xi_2 \\
    L_{ji} &= \intxip\intxit 
    \mathcal{G}(\xi_2,\xi_3;\xi_{2,j},\xi_{3,j})
    ~d\xi_3 d\xi_2,
\end{align*}
where $\xi_{2,j}$ and $\xi_{3,j}$ are fixed at the external face center of cell $j$. The precise forms of the integrands depend on whether one is working in axial symmetry or not, and we give explicit forms for both in the following.

\subsection{\label{sec:numerical-method-2d}2D axisymmetric solutions}
Assuming both geometry and solutions are axisymmetric, the matrix elements are given by:
%\begin{widetext}
\begin{align}
    {K_{ji}}^{\mathrm{sym}} &= 
    \intxip
    \left[
    \dfrac{\mu^3}{8\pi\sqrt{RR_j}}
    \dfrac{E(\mu)}{(1-\mu^2)}J\bm{n}
    \cdot\left(\dfrac{\partial\chi}{\partial Z}\ez - 
    \dfrac{\partial\chi}{\partial R}\er\right)
    + \dfrac{\mu K(\mu)}{4R^\threehalf\pi\sqrt{R_j}}
    J\bm{n}\cdot\er  
    \right]
    d\xi_2
    \\
    {L_{ji}}^{\mathrm{sym}} &= -\intxip \dfrac{\mu K(\mu)}{2\pi\sqrt{RR_j}} ~d\xi_2,
\end{align}
%\end{widetext}
where
\begin{align}
    \chi &= \chi(R,Z;R',Z') = \dfrac{R^2 + (R')^2 + (Z-Z')^2}{2RR'}, \\
    \mu^2 &= \dfrac{2}{\chi+1},
\end{align}
and
\begin{align*}
    R = R(\xi_2), \quad R_j = R(\xi_{2,j}),
    \quad Z = Z(\xi_2),\quad
    \mu = \mu(\xi_2,\xi_{2,j}), \\ 
    \chi = \chi(\xi_2,\xi_{2,j}),\quad
    J=J(\xi_2),\quad 
    \bm{n} = \bm{n}(\xi_2),
\end{align*}
and $K(\mu)$ and $E(\mu)$ are complete elliptic integrals of the first and second kinds, respectively (see~\ref{sec:elliptic}).
When the source and the target points $i$ and $j$ are co-located ($i=j$), $\mu=1$ at the center of the integral interval and the elliptic integral $K(\mu)$ diverges. The term in $E(\mu)$ in the integrand of $K_{ij}$ also diverges due to division by $(1-\mu^2)$. Therefore, the integrands of both $L_{ij}$ and $K_{ij}$ diverge at $\mu=1$ and we must use a specialized quadrature rule as described in  \ref{sec:quadrature}.

\subsection{\label{sec:numerical-method-3d} 3D solutions}
In 3D (but still axisymmetric geometry), the matrix elements take the form
\begin{align*}
    K_{ji} =
    -
    \intxip
    \dfrac{J}{8\pi \sqrt{2RR_j}}
    \left\{
    \dfrac{\n\cdot\er}{R}
    \intxit
    \dfrac{1}{\sqrt{\chi-\cos(\xi_3-\xi_{3,j})}}
    d\xi_3
    + \right.\\
    \left.
    (\n\cdot\er\frac{\partial\chi}{\partial R} + 
    \n\cdot\ez\frac{\partial\chi}{\partial Z})
    \intxit 
    \dfrac{1}{(\chi-\cos(\xi_3-\xi_{3,j}))^\threehalf}
    d\xi_3 \right\} d\xi_2
    \\
    L_{ji} = \intxip
    \dfrac{1}{4\pi\sqrt{2RR_j}}
    \intxit
    \dfrac{1}{\sqrt{\chi-\cos(\xi_3-\xi_{3,j})}}d\xi_3 d\xi_2,
\end{align*}
which contain integrands that diverge when $i=j$.
The troublesome integrals have the form
\begin{align*}
    \intxit
    \dfrac{d\xi_3}{
    (\chi-\cos(\xi_3-\xi_{3,j}))^\alpha}
\end{align*}
where $\alpha=\half$ for $L_{ji}$ and $\alpha=\threehalf$ for $K_{ji}$. First we consider the case with $\alpha=\half$. This integral can be expressed in the following way,
\begin{align*}
    &
    \intxit
    \dfrac{d\xi_3}{
    (\chi-\cos(\xi_3-\xi_{3,j}))^\half}
    =
    \dfrac{2}{\sqrt{\chi+1}}
    \left[
    F(\beta^+|\mu)
    -
    F(\beta^-|\mu)
    \right],
\end{align*}
where
\begin{align}
\beta^\pm = \dfrac{\xi_{3,i}-\xi_{3,j}+\pi}{2} \pm \dfrac{\Delta\xi_3}{4}.
\end{align}
The function $F(\gamma|\mu)$ is the incomplete elliptic integral of the first kind where $\gamma$ is the amplitude and $\mu$ is the modulus.
Beginning in a similar manner as above, the integral with $\alpha=\threehalf$ appearing in $K_{ij}$ can be expressed as follows
\begin{align*}
    &
    \intxit
    \dfrac{d\xi_3}{
    (\chi-\cos(\xi_3-\xi_{3,j}))^\threehalf}
    = 
    \dfrac{2}{(\chi+1)^\threehalf}
    \int_{\beta^-}
    ^{\beta^+}
    \dfrac{d\theta}
    {(1-\mu^2\sin^2\theta)^\threehalf},\\
    \vspace{10pt}
\end{align*}
where $\theta = \xi_3 - \xi_{3,j}$.
This contains an incomplete elliptic integral of the third kind (see~\ref{sec:elliptic}),
with $n=\mu^2$ and $m=\mu$, and where $x=\sin\Omega$.
Therefore, we finally obtain for $L_{ij}$ and $K_{ij}$ in 3D axisymmetric geometries:
\begin{align*}
    {K_{ji}}^\mathrm{asym} &=
    -
    \intxip
    \dfrac{J}{4\pi\sqrt{2RR_j}}
    \left\{
    \dfrac{\n\cdot\er}{R\sqrt{\chi+1}}
    [ F(\beta^+ |\mu) - F(\beta^-|\mu)]
    \right.
    \\
    &
    \left.
    +
    \dfrac{1}{(\chi+1)^\threehalf}
    \left(\n\cdot\er\dfrac{\partial\chi}{\partial R} + 
    \n\cdot\ez\dfrac{\partial\chi}{\partial Z}\right)
    [
    \Pi(\beta^+|\mu^2,\mu)
    - \Pi(\beta^-|\mu^2,\mu)
    ]
    \right\}
    d\xi_2, \\
    {L_{ji}}^\mathrm{asym} &= 
    \intxip
    \dfrac{1}{2\pi\sqrt{2RR_j(\chi+1)}}
    [ F(\beta^+ |\mu) - F(\beta^-|\mu)]
    d\xi_2,
\end{align*}
These integrands still contain singularities at $i=j~(\mu=1)$, and we use a specially formulated quadrature in $\xi_2$ as discussed in  \ref{sec:quadrature}.

\subsection{Solution strategy}
The $K_{ji}$ and $L_{ji}$ matrices are dense, with their size scaling as $N^2$ in 2D and $N^4$ in 3D, where $N$ is the number of points in one angular dimension in the toroidal surface. The cost of inverting $K_{ij}$ scales as $N^3$ in 2D and and $N^6$ in 3D. Therefore, the cost of calculating the matrix $K_{ji}^{-1}L_{ji}$ may become prohibitively large for large $N$, particularly in 3D (e.g., increasing $N$ by a factor of 2 may increase the computational complexity in 3D by up to a factor of $64$).
That said, since the matrices $K_{ji}$ and $L_{ji}$ contain only geometric terms and the mesh in PIXIE3D is static, calculating the matrix elements and inverting $K_{ji}$ need only be performed once during initialization and can then be stored, which enables amortization of the setup costs during the lifetime of the simulation.

In our PIXIE3D implementation, we rely on the numerical software library PETSc \citep{petsc-user-ref,petsc-efficient} to compute the matrix elements and invert the corresponding matrices in parallel. In particular, we use PETSc's interfaces to the ELEMENTAL software package \citep{Elemental2012}, which supports dense-matrix parallel linear algebra. Our numerical experiments will show competitive performance even with fairly refined angular meshes in 3D, of order a few minutes for a $128\times128$ angular mesh in 8 ranks distributed over the surface. Larger meshes will incur longer setup times, and may require additional strategies for cost control. One possible strategy to explore in the future is the use of a high-order discretization of the BIM integrals, which could afford a significant decrease of required number of mesh points for a target error level.

\section{Extension of the BIM module to include a external perfectly conducting wall}
\label{sec:external_wall}

Many Tokamak designs consist of a secondary wall exterior to the resistive plasma-facing wall and separated by a so-called `blanket` region. We consider the effects of this wall by extending our treatment to include a second exterior perfectly conducting wall outside of the resistive wall, between which we assume there is vacuum. Since the external wall is assumed to be perfectly conducting, $\partial_t B^n=0$ and the wall maintains its nascent normal magnetic field (or equivalently, it conserved the poloidal flux) from the time that the magnetohydrostatic equilibrium is established. 

In axisymmetric cylindrical geometry, the external wall is characterized by a parametrization of its coordinates $(R,Z)$ and the equilibrium poloidal flux $\Psi$ in terms of PIXIE3D's poloidal angular coordinate, $\xi_2$. The poloidal flux function uniquely determines the normal magnetic-field component at the wall (as we show below), and is usually found from an external Grad-Shafranov solver such as EFIT. It does not evolve in time owing to the perfect-conducting property of the external wall. 
The wall surface coordinates and corresponding poloidal flux are discretized independently from PIXIE3D as a collection of points. 
The linear system (\ref{eq:linear_system}) is extended to include this external wall mesh, and we solve concurrently for the vacuum potential on both walls.

We need to provide two quantities at the external-wall mesh points to augment the external-wall BIM treatment: the outward-facing surface normal vector $\mathbf{n}_{ext}$(\ref{eq:pixie3d-normal})
and the surface normal component of the magnetic field $B_{ext}^{n}$~(\ref{eq:bn-definition}). These can be readily found from the parameterizations above. In particular, 
from (\ref{eq:pixie3d-normal}), the normal vector $\mathbf{n}$ is given as:
\begin{equation}
    \mathbf{n}_{ext} = \left . \frac{1}{J}\left(\frac{\partial\mathbf{r}}{\partial\xi_{2}}\times\frac{\partial\mathbf{r}}{\partial\xi_3}\right) \right |_{ext}.
\end{equation}
Here, $\xi_3$ is the toroidal angle, which is the negative of the cylindrical azimuthal angle, $\xi_3=-\phi$. Since the surface position is parametrized with $\xi_2$ $\mathbf{r}=R(\xi_{2})\er+Z(\xi_{2})\ez$, we find:
\[
\frac{\partial\mathbf{r}}{\partial\xi_3}=-\ephi R\,\,;\,\,\frac{\partial\mathbf{r}}{\partial\xi_{2}}=\frac{\partial R}{\partial\xi_{2}}\er+\frac{\partial Z}{\partial\xi_{2}}\ez,
\]
and therefore (since ($R,\phi,Z)$ is right-handed):
\[
J\mathbf{n}_{ext}=-R\left[\frac{\partial R}{\partial\xi_{2}}\er+\frac{\partial Z}{\partial\xi_{2}}\ez\right]_{ext}\times\ephi= \left [ R\frac{\partial Z}{\partial\xi_{2}}\er-R\frac{\partial R}{\partial\xi_{2}}\ez \right ]_{ext}.
\]
Since the parametrization of the external-wall coordinates may have arbitrary orientation, clockwise
or counter-clockwise, it is important to check that the normal vector is pointing outwards of the domain of interest. To ensure an outward-pointing
normal, we check if a surface point augmented with the normal is enclosed
by the external-wall surface or not, i.e., whether $\mathbf{r}_{ew}+\epsilon J\mathbf{n}\in\Omega_{ew}$
or not, with $\epsilon$ a small number. If the point is enclosed
by the surface, then the normal is pointing inward, and we need to
flip the sign. 

The normal component of the magnetic field $B^n$ is found from the definition of the normal above as:
\[
B_{ext}^{n}= J \mathbf{n}\cdot\mathbf{B}|_{ext}=\left [ \mathbf{B}\cdot\left(\frac{\partial\mathbf{r}}{\partial\xi_{2}}\times\frac{\partial\mathbf{r}}{\partial\xi_3}\right) \right]_{ext}.
\]
The magnetic field is usually given in a mixed representation in terms of
the poloidal flux $\Psi$ and the toroidal magnetic field $B_{\xi_3}$ as:
\[
\mathbf{B}=\nabla\Psi\times\nabla\xi_3+RB_{\xi_3}\nabla\xi_3,
\]
where $\nabla\xi_3=-\ephi/R$. It follows that:
\begin{align}
\label{eq:b_n_ext}
B_{ext}^{n}=\left(\frac{\partial\mathbf{r}}{\partial\xi_{2}}\times R\ephi\right)_{ext}\cdot\left(\nabla\Psi\times\frac{\ephi}{R}\right)_{ext}=\left[\nabla\Psi\cdot \left( \ephi\times\left(\frac{\partial\mathbf{r}}{\partial\xi_{2}}\times\ephi\right)\right) \right]_{ext}
=\left. \nabla\Psi\cdot\frac{\partial\mathbf{r}}{\partial\xi_{2}}\right|_{ext}=\left. \frac{\partial\Psi}{\partial\xi_{2}}\right|_{ext},
\end{align}
where we have used that $\dfrac{\partial\mathbf{r}}{\partial\xi_{2}}\cdot\ephi=0$
in axisymmetric geometries.

\section{Extension to include external coils}
\label{sec:external_coils}
External coils carry current that modify the magnetic field and can be used as an additional means to control the plasma in the tokamak. Their influence can be modelled as a modification of the vacuum magnetic field exterior to the wall of the tokamak, and included in the boundary conditions for the MHD PDE system. We describe this process next.

The vacuum magnetic field generated by an infinitesimally thin filament $i$ is given by the Biot-Savart law as
\begin{align*}
    \B_i(\rr) = \dfrac{\mu_0}{4\pi}
    \int_C \dfrac{I_id\lb_i\times\rp_i}{|\rp_i|^3},
\end{align*}
where $\rp_i = \rr - \lb_i$ is the displacement vector of the point $\rr$ in space from the line segment $d\lb$ and $I$ is the current in the filament in the direction of vector $d\lb$. Stationary flow is assumed within the filament meaning that divergence of current density $\nabla\cdot \mathbf{j}=0$ and therefore I is uniform within the filament. Subscript $C$ denotes the integral over the current-carrying circuit. The total vacuum magnetic field from $i=1,\dots,n$~filaments is obtained using superposition. Assuming that the current carried by a filament is constant along the circuit,
\begin{align*}
    \B(\rr) = \dfrac{\mu_0}{4\pi}
    \sum_{i=1}^n I_i \int_C
    \dfrac{d\lb_i\times\rp_i}
    {|\rp_i|^3},
\end{align*}
and we can therefore represent an external coil as a discrete collection of filaments distributed in space. A full derivation of the expression for the resulting field for an axisymmetric filament in cylindrical coordinates is provided in~\ref{sec:biot-savart}.

With the coils present, the magnetic field external to the resistive wall is a superposition of two fields: a singular field generated by the coils and a regular field generated by vacuum response to the plasma and resistive wall:
\begin{align}
\label{eq:btotal-coils}
    \B = \B_\mathrm{coils} + \B_\mathrm{vacuum},
\end{align}
and because the solenoidal constraint requires that the normal magnetic field is continuous at the resistive wall, this means
\begin{align}
    B^n_\mathrm{vacuum} = B^n_\mathrm{plasma} -
    B^n_\mathrm{coils},
\end{align}
where $B^n_\mathrm{plasma}$ is an evolved quantity and $B^n_\mathrm{coils}$ is given by the Biot-Savart law. The regular field from the vacuum response can then be calculated from the corresponding scalar potential solution using the BIM with right hand side $B^n_\mathrm{vacuum}$, giving $\B_\mathrm{vacuum}$ and therefore the total $\B$ external to the resistive wall from \ref{eq:btotal-coils}. The jump in the tangential field can therefore be computed to give the wall electric field and this completes the external wall treatment.

\section{Numerical verification of BIM solver in PIXIE3D}
\label{sec:verification}

We introduce next several analytical verification tests that can be implemented by any practitioner to test their MHD RW boundary treatment. We begin by proposing a toroidal infinite-domain vacuum test, based on an analytical solution to the vacuum Laplace equation in the external region of a circular cross-section torus is used. We follow with a toroidal finite-domain vacuum test, where we consider the analytical solution of a cylindrical tokamak as the limit of a very large aspect-ratio torus. We then consider a test for the treatment of external coils. This test is less accessible, as it requires a more realistic setup, but we provide enough information about wall configuration and boundary conditions for any practitioner to reconstruct the test without additional input.

\subsection{Toroidal infinite vacuum domain}

\begin{figure*}[t]
    \centering
    \includegraphics[width=\textwidth]{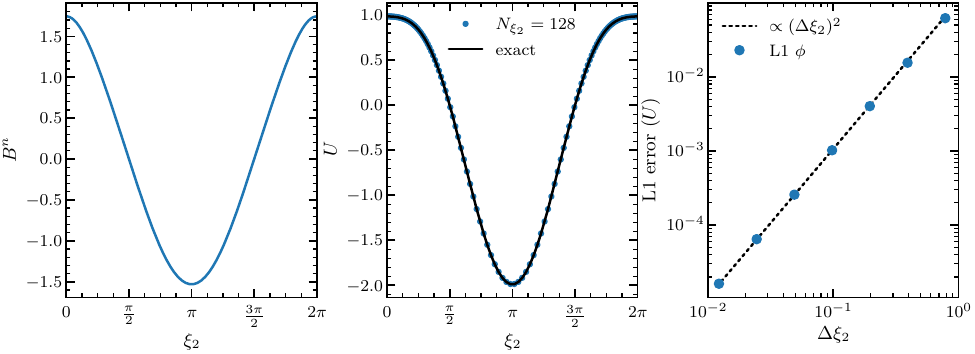}
    \caption{Verification results for the 2D axisymmetric toroidal infinite vacuum analytical test. Normal magnetic field (left panel), solution for the magnetic scalar potential $U$ with analytic solution (middle panel) and second order convergence to analytic solution (right panel). $B^n$ is given by a toroidal harmonic with $n=0$ and $m=1$.}
    \label{fig:toroidal_solution}
\end{figure*}

\begin{figure*}
    \centering
    \includegraphics[width=\linewidth]{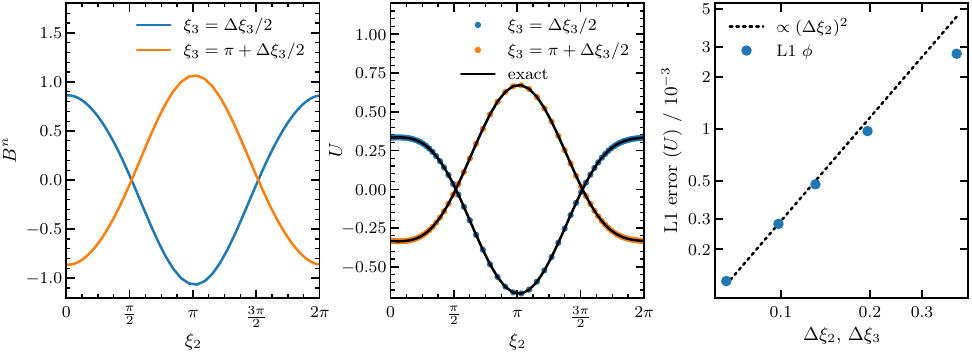}
    \caption{Verification results for the 3D toroidal infinite vacuum analytical test. Normal magnetic field (left panel), solution for the magnetic scalar potential $U$ with analytic solution (middle panel) and second order convergence to analytic solution (right panel). $B^n$ is given by a toroidal harmonic with $n=1$ and $m=1$.}
    \label{fig:PIXIE3d_3dconv}
\end{figure*}

In order to obtain an analytic solution to verify our methodology, we consider a toroidal coordinate system $\xtor = \xtor(\rho,\theta,\phitor)$ by its relation to the Cartesian coordinate system $\x=\x(x,y,z)$
\begin{align*}
    x = a\dfrac{\sinh{\rho}\cos\phitor}{\cosh{\rho} - \cos\theta}, \\
    y = a\dfrac{\sinh\rho\sin\phitor}{\cosh\rho-\cos\theta}, \\
    z = a\dfrac{\sin\theta}{\cosh\rho-\cos\theta},
\end{align*}
and to the cylindrical coordinate system $\bm{x}_\mathrm{cyl}(R,\phi,Z)$ by
\begin{align*}
    R &= a\dfrac{\sinh\rho}{\cosh\rho-\cos\theta}, \\
    Z &= a\dfrac{\sin\theta}{\cosh\rho-\cos\theta},\nonumber\\
    \phi &= -\phitor.
\end{align*}
giving, for the toroidal coordinate system, Laplace's equation in the form \cite{lebedev1972a}
\begin{align}
    \label{eq:laplace_toroidal}
    \dfrac{(\cos\theta - \cosh{\rho})^3}{\sinh{\rho}}
    \left[
    \dfrac{\partial}{\partial\rho}\left(\dfrac{\sinh\rho}{\cosh\rho-\cos\theta}\dfrac{\partial U}{\partial\rho}\right) + 
    \right.\nonumber\\
    \left.
    \dfrac{\partial}{\partial\theta}\left(\dfrac{\sinh\rho}{\cosh\rho-\cos\theta}\dfrac{\partial U}{\partial\theta}\right) + 
    \right.\nonumber\\
    \left.
    \dfrac{1}{(\cosh\rho-\cos\theta)\sinh\rho}\dfrac{\partial^2U}{\partial\phitor^2}\right] = 0.
\end{align}
This is made separable using the ansatz
\begin{align}
    \nonumber
    U = v\sqrt{2\cosh\rho-2\cos\theta}; \quad 
    v = A(\rho) B(\theta) U(\phi),
\end{align}
yielding the general solution %\swj{potential should be V to be consistent with the rest of the paper.}
\begin{align}
    \label{eq:general_toroidal_laplace}
    U = \sqrt{2z - 2\cos\theta}
    \sum_{m=0}^\infty
    \sum_{n=0}^\infty
    \left[
    \alpha^P_{mn}P^n_{m-\half}(z)
    +
    %\right.\nonumber\\
    %\left.
    \alpha^Q_{mn}Q^n_{m-\half}(z)
    \right]
    A_{m,n} e^{i(m\theta+n\phi_\mathrm{tor})} ,
    n\in\mathbb{Z}_0^+, m\in\mathbb{Z}_0^+,
\end{align}
where $z=\cosh\rho$.
%In (\ref{eq:laplace_toroidal_general}), 
$P_{m-\half}^n$ and $Q_{m-\half}^n$ are associated Legendre functions of the first and second kind of order $n$ and degree $m-\half$, also known as toroidal functions or toroidal harmonics. The integers $m$ and $n$ are the poloidal and toroidal mode numbers, respectively.
$A_{m,n}$ is the amplitude of mode $(m,n)$ and $\alpha^P_{mn}$ and $\alpha^Q_{mn}$ are constants determined from the choice of boundary conditions.
For the semi-infinite domain we are considering, the coefficient of $Q^n_{m-\half}$ must equal zero ($\alpha^Q_{mn}=0$) if the solution is regular, because as the distance from the toroidal surface tends to infinity,
\begin{align*}
    \lim_{z\rightarrow 1^+}Q_\nu(z) = \infty.
\end{align*}
Therefore, in the semi-infinite domain the solution for a single mode is
\begin{align}
    \label{eq:laplace_toroidal_single}
    U_{m,n} = \sqrt{2z - 2\cos\theta}
    \alpha_{mn}P^n_{m-\half}(z)
    e^{i(m\theta+n\phi_\mathrm{tor})} ,
    n\in\mathbb{Z}_0^+, m\in\mathbb{Z}_0^+,
\end{align}
where we have absorbed the mode amplitude into the constant $\alpha$.
This corresponds to a normal magnetic field of
\begin{align}
    B^\rho = \frac{\alpha \sqrt{z -\cos \theta } e^{i (m\theta  +n \phi_\mathrm{tor} )}}{2 a\sinh\rho}
    \left\{[z  (\cos \theta +2 m \cos \theta -2 m 
    z ) 
    - 1]P_{m-\half}^n(z ) 
    \right.\\
    \left.
    - (2 m-2 n+1) (\cos \theta -z ) P_{m+\half}^n(z )\right\},
\end{align}
which we use as input to our method in order to find the potential $U_{m,n}$ to verify against the solution (\ref{eq:laplace_toroidal_single}).
Defining the major and minor radii of the toroidal plasma wall as $R=1.65$ and $r=1$, respectively, the parameters are
\begin{align}
    \alpha &= 1, \\
    \rho &= \operatorname{arcosh}\left(
    \dfrac{R}{r}
    \right), \\
    a &= \sqrt{
    \dfrac{R + r}{R - r}
    }.
\end{align}
In Figure~\ref{fig:toroidal_solution} we show the agreement between the analytic solution and the axisymmetric numerical solution, which converges at second order, as expected. Similar convergence is shown in Figure~\ref{fig:PIXIE3d_3dconv} for 3D.

We use this toroidal example to exemplify performance of our BIM implementation. Table \ref{tab:scaling} shows timings in seconds for combined matrix setup and inversion operations for the 3D BIM module. These results indicate that BIM setup costs are manageable (of order a few minutes for the finest $128\times128$ angular mesh with 8 ranks distributed over the surface), are dominated by the matrix setup stage with the wall-clock time scaling as $\mathcal{O}(N^4)$ instead of $\mathcal{O}(N^6)$, and scale reasonably well in parallel (with 80-90\% strong-scaling efficiency from 1 to 8 ranks). 

\begin{table}[t]
\centering
\caption{Performance (wall-clock) timings and parallel scaling results of the 3D BIM module for a circular torus for various grid resolutions in the angular coordinates $(\xi_2,\xi_3)$. Timings are obtained in a 2024 MacBook Pro with Apple Silicon processors.}
\begin{tabular}{lcccc}
\hline
{$(\xi_2,\xi_3)$-grid} & {1 rank} & {8 ranks} & {Parallel efficiency} & {CPU scaling $N^{\alpha}$} \\
\hline
16x16   & 0.3s   & 0.05s & 0.8 & {} \\
32x32   & 4.5s   & 0.68s  & 0.8 & $\alpha=$3.7 \\
64x64   & 75s    & 11.7s  & 0.8 & $\alpha=$4.1 \\
128x128 & 1250s  & 223s   & 0.7 & $\alpha=$4.2 \\
\hline
\end{tabular}
\label{tab:scaling}
\end{table}


\subsection{\label{sec:toroidal_finite}Toroidal finite vacuum domain}
We are now interested in the solution in a finite domain enclosed by a resistive inner and a perfectly conducting outer wall at $\rho=\rho_r$ and $\rho=\rho_c$, respectively, with minor radii $r_r < r_c$ and therefore $\rho_r > \rho_c$. 
The general solution to Laplace's equation in toroidal coordinates is given in (\ref{eq:general_toroidal_laplace}).
Because of the geometric factor appearing in front of the summation, the solution to (\ref{eq:general_toroidal_laplace}) cannot be obtained trivially as in the case of cylindrical or spherical coordinate systems to find the constants $\alpha^P_{mn}$ and $\alpha^Q_{mn}$ for a given set of boundary conditions \cite{kuyucak1998a}.
Instead, we use here the solution for an infinite cylinder and take the limit of large major radius of the torus to approximate the infinite cylinder.

Following \cite{schnack1990computational,spinicci2023a}, we obtain that for an infinite cylinder and a cylindrical mode with mode number $m$, the solution for the potential is given by
\begin{align*}
    U_m(r,\theta) = \dfrac{A_m}{m}\dfrac{r_c^m}{1-r_c^{2m}}
    \left[
    \left(\dfrac{r}{r_c}\right)^m + \left(\dfrac{r_c}{r}\right)^m
    \right]
    e^{im\theta}
\end{align*}
where $A_m$ is the Fourier amplitude of mode $m$ appearing in the boundary condition at the inner resistive wall surface ($r = r_r = 1$) for the normal magnetic field:
\begin{align*}
    B^n|_{r=1} = \partial_rU_m = 
    \dfrac{A_m}{m} \dfrac{r_c^m}{1-r_c^{2m}}
    \left( \dfrac{m}{r_c^m} - mr_c^m \right) e^{im\theta},
\end{align*}
and $r_c$ is the radius of the outer conducting wall at which the boundary condition $B^n|_{r=r_c}$ is imposed.
We note that care must be taken to use the correct sign for $B^n$ commensurate with the direction of the normal vector on the respective surface. Furthermore, in our case when approximating the infinite cylinder with a torus of large major radius, $B^n$ should be multiplied with the Jacobian $J=Rr$.
\begin{figure*}
    \centering
    \includegraphics[width=\textwidth]{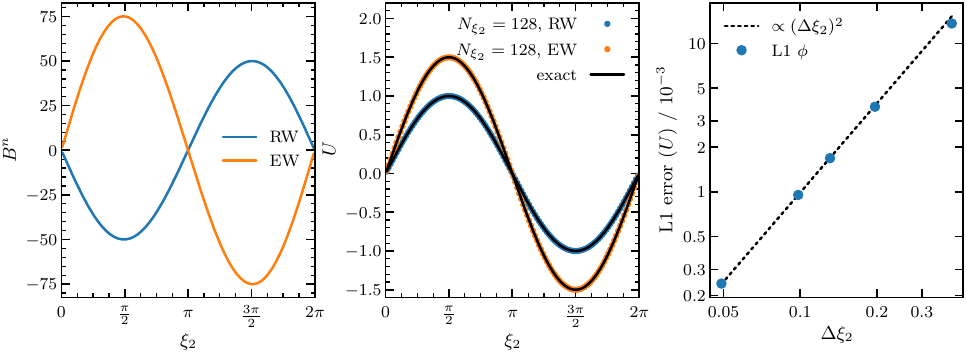}
    \caption{Verification results for 2D infinite cylinder double-wall problem. \emph{Left:} Normal component of magnetic field on the two surfaces for the $m=1$ mode problem. RW is the resistive wall and EW is the external wall. \emph{Middle:} Comparison of analytic and numerical solutions. \emph{Right:} Convergence analysis.}
    \label{fig:ext_soln_m1_2d}
\end{figure*}
\begin{figure}
    \centering
    \includegraphics[]{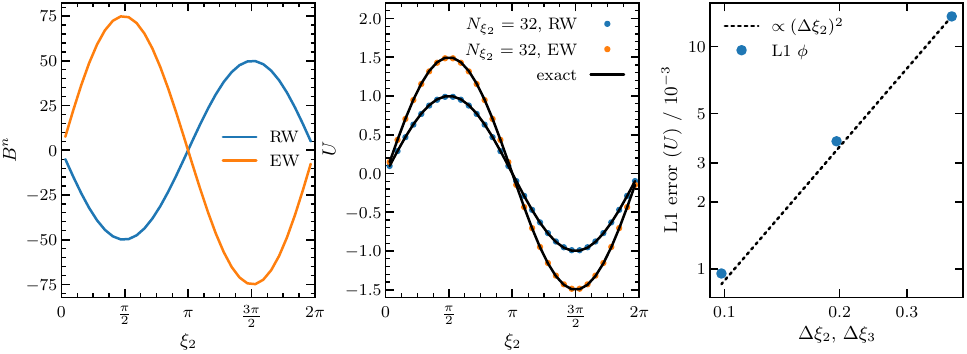}
    \caption{Verification results for 3D infinite cylinder double-wall problem. \emph{Left:} Normal component of magnetic field on the two surfaces for the $m=1$ mode problem at $\xi_3=\Delta\xi_3/2$. RW is the resistive wall and EW is the external wall. \emph{Middle:} Comparison of analytic and numerical solutions at $\xi_3=\Delta\xi_3/2$. \emph{Right:} Convergence analysis.}
    \label{fig:ext_soln_m1_3d}
\end{figure}
If we have zero normal magnetic field at the resistive plasma-facing wall $B^n|_{r=r_r}=0$ and non-zero normal field at the conductive wall $B^n|_{r=r_c}$
the potential is given by
\begin{align*}
    U_m(r,\theta) = \dfrac{A_m}{m}\dfrac{r^m + r^{-m}}{r_c^{m-1}-r_c^{-(m+1)}}
    e^{im\theta}.
\end{align*}
If both walls have non-zero $B^n$, then the potential is
\begin{align*}
    U_m(r,\theta) =
    \frac{A_m^c r_c^{m+1} \left(r^{2 m}+1\right)-A_m^r \left(r_c^{2 m}+r^{2 m}\right)}{r^m m \left(r_c^{2 m}-1\right)}
    e^{im\theta}
\end{align*}
where $A_m^r$ and $A_m^c$ are the Fourier amplitudes at the resistive and conductive walls, respectively.
For this test, we use $r_r=1$, $r_c=1.5$, $A_m^c=A_m^r=1$, a major radius of $R=50$ and a mode number of $m=1$. 
The result for this case is shown in Figure~\ref{fig:ext_soln_m1_2d} for the 2D solution and in Figure~\ref{fig:ext_soln_m1_3d} for the 3D solution.
We note that if the distance separating the resistive and conducting walls is on the order of or less than the grid spacing, the quality of the solution is severely affected and therefore such configurations should be avoided.

\subsection{\label{sec:external_coils_verif}External coil treatment}
It is challenging to devise a method for the direct verification of the external coil treatment. Instead, we resort to a comparison of the tangential magnetic field on the vacuum side of the resistive wall calculated using (1) the verified external wall treatment and (2) the external coil treatment. Our setup is an NSTX EFIT equilibrium and corresponding coil configuration from \cite{krebs2020a} (see supplemental material, with coil currents in MA instead of kA; the latter is a typo in the reference). When using the external wall, the normal magnetic field at the external wall $B^n_\mathrm{ext}$ (needed to find the BIM vacuum solution) is computed according to (\ref{eq:b_n_ext}) from the equilibrium flux field provided in the NSTX EFIT equilibrium file, which includes the response to the external coils. When using the external coils, we simply use the coil configuration that was used to derive the equilibrium in the first place. Both methods should result in very similar tangential magnetic fields on the vacuum side of the resistive wall. All of the data required to reproduce the setup for this verification test is provided in the supplemental material, in file \texttt{rw\_nstx\_all\_data.txt}.

\begin{figure}[t]
    \centering
    \includegraphics[]{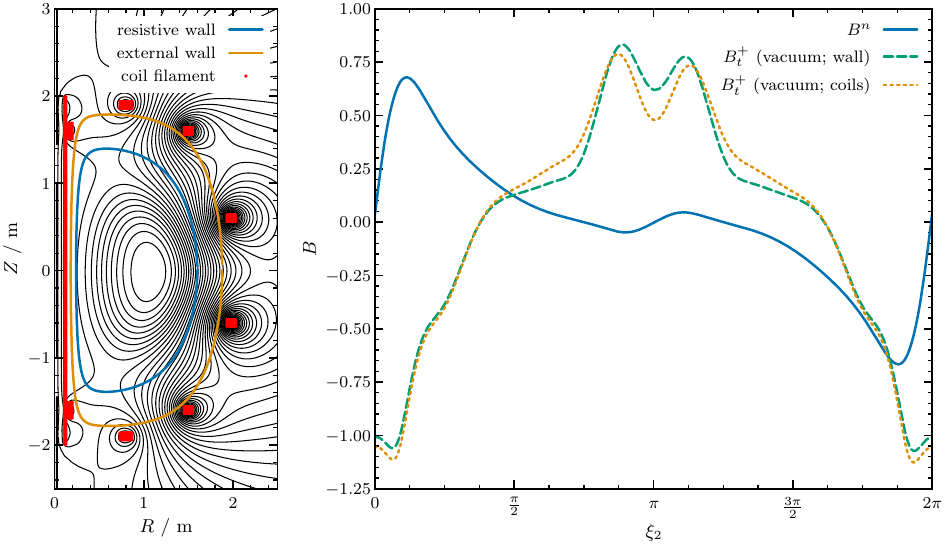}
    \caption{Verification of external coil treatment against external wall approximation for the NSTX equilibrium from \cite{krebs2020a}. Left panel: equilibrium poloidal flux generated from the coils and plasma current, coil locations (red), PIXIE3D's resistive wall (blue) and the external wall location used for the verification (orange). Right panel: good qualitative agreement is observed in the vacuum B-tangential at the resistive wall ($B_t^+$) between the external coil and external wall vacuum-field treatments. The normal component $B^n$ at the resistive wall (which is the input to the vacuum BIM solver) is also provided for verification purposes.
    }
    \label{fig:nstx-coil-verification}
\end{figure}

The left panel in Figure~\ref{fig:nstx-coil-verification} shows the poloidal field shape of the equilibrium, coil locations (red), PIXIE3D's resistive wall (blue) and the external wall when used (orange). There are caveats to this verification approach. Firstly, we know that the external wall treatment breaks down if the external ideal and plasma resistive walls are too close together (see Section~\ref{sec:external_wall}). Secondly, the external wall treatment approximates the singular coil field response by a regularized vacuum field. This approximation worsens the closer the external wall is to the coils. The error incurred in both cases will be more prominent close to the $R=0$ symmetry axis, where the coils, external and resistive walls are closest together.
The right panel in Figure~\ref{fig:nstx-coil-verification} shows the resulting tangential magnetic field calculated using both vacuum treatments. Indeed, they do qualitatively agree, and as expected the error is largest around $\theta=\pi$, close to the symmetry axis at $R=0$. We interpret this result as confirmation that the external coil treatment is correctly implemented.

\section{\label{sec:VDE-ITER}Application: Vertical displacement event (VDE) in ITER}

\begin{figure}[ht!]
    \centering
    \includegraphics[scale=0.25]{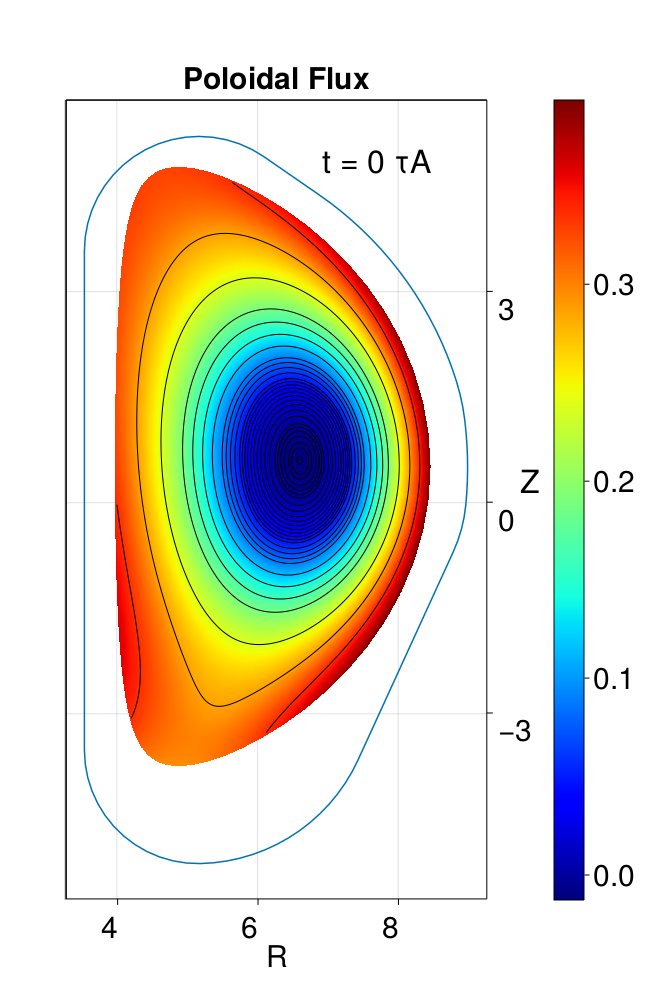}
    ~\includegraphics[scale=0.25]{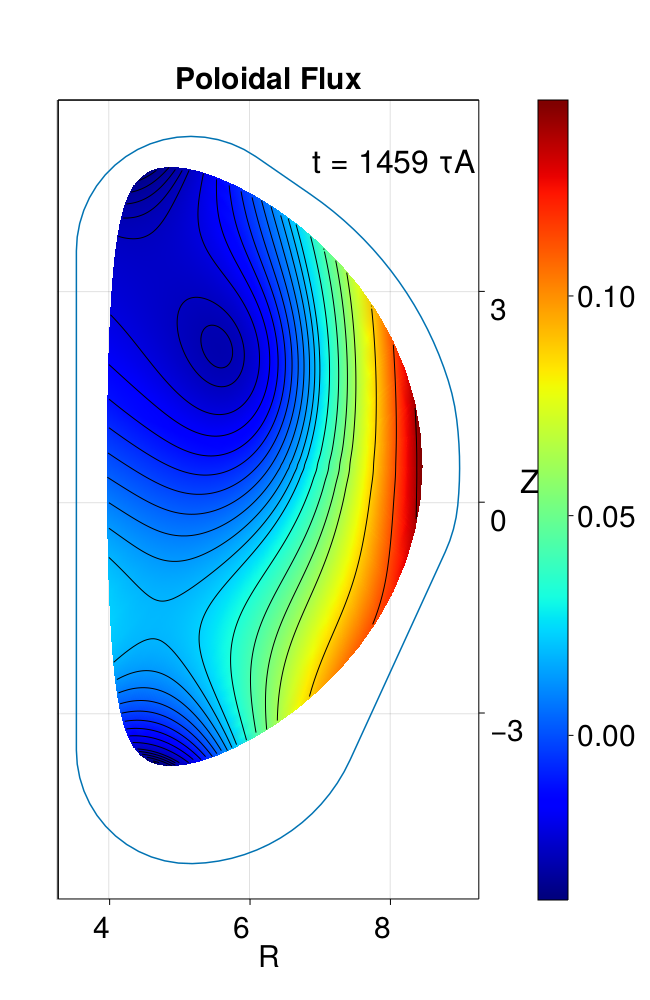}\\
    \includegraphics[scale=0.25]{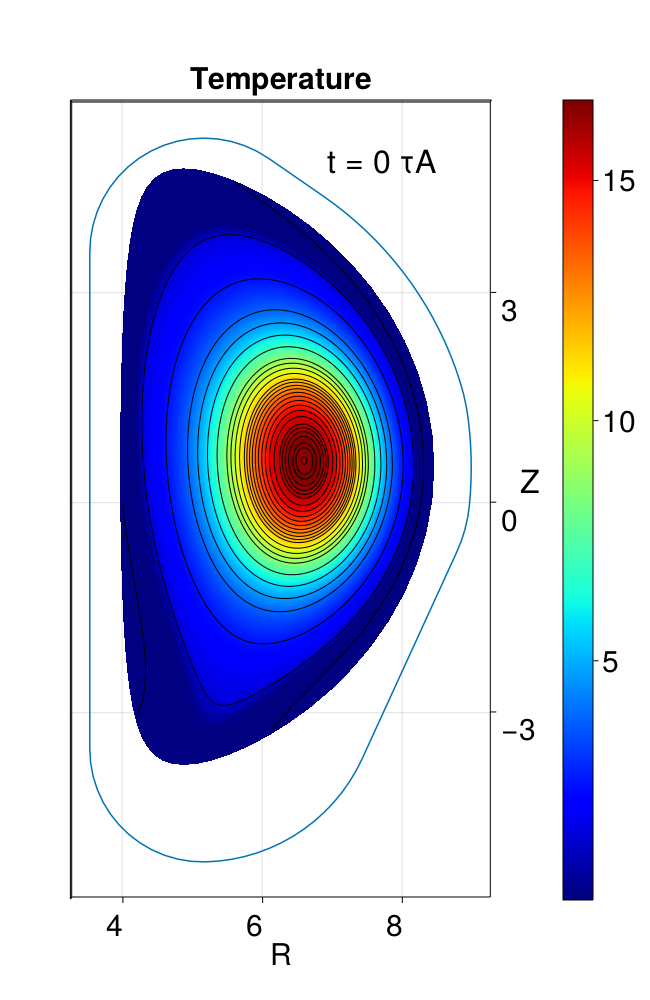}
    ~\includegraphics[scale=0.25]{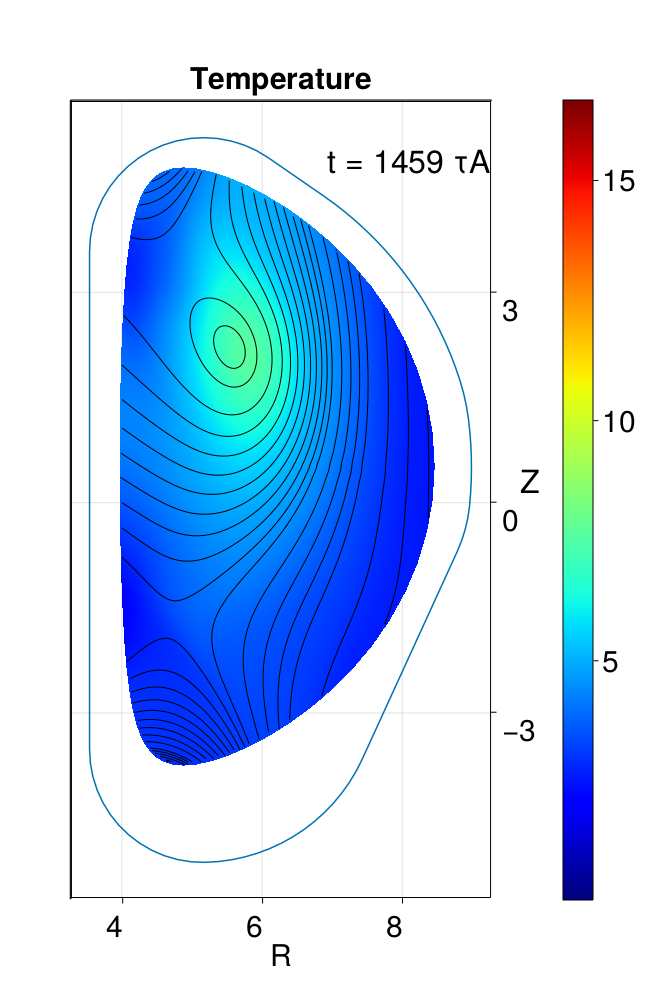}
    \caption{The poloidal flux and temperature (keV) for a PIXIE3D thin resistive wall simulation of a VDE in ITER with a resistive wall time $\tau_{w} = 20 \tau_{A}$ and resistivity $S^{-1} = 10^{-4}$. Plots on the left are the initial conditions, on the right is during the VDE. The simulations continue until the magnetic axis has completely moved into the vacuum region. The geometry of the perfectly conducting external wall is shown in blue. Flux surfaces are shown in black in all plots.}
    \label{fig:pixieVDE}
\end{figure}

We demonstrate our PIXIE3D implementation of the RW boundary treatment with the simulation of a vertical displacement event (VDE) in ITER. The initial condition is a 9~MA ITER equilibrium for which we provide the $q$-profile in the supplemental material. The full EFIT equilibrium file is available upon request. We consider the plasma to be limited by a thin resistive wall, with a perfectly conducting external wall surrounding the vacuum chamber (as detailed in Fig. \ref{fig:pixieVDE}).
We consider a resistive wall time $\tau_{w} = 20 \tau_{A}$ and normalized plasma resistivity $\eta = 10^{-4}$ and viscosity $\mu=10^{-3}$. Thermal diffusivity coefficients (dimensionless) are $\chi_{\perp} = 10^{-5}$, $\chi_{\parallel} = 10$. Results are shown in Figure \ref{fig:pixieVDE}.
The resolution in logical coordinates $(r,\theta,\phi)$ is $128\times64\times1$.
A VDE is seen to take place on a timescale of $\sim10^3 \tau_A$, where $\tau_A$ is the Alfv\'en time. The strong parallel heat transport results in good energy confinement in the core early on, and to a thermal quench once the separatrix intercepts the first wall.

\begin{figure}[ht!]
    \centering
    \includegraphics[scale=0.5]{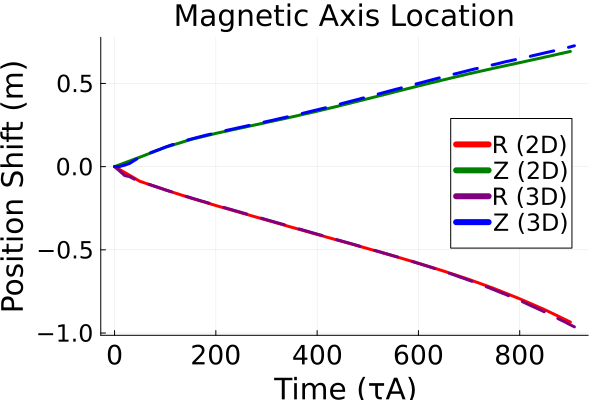}
    \caption{Shift of the magnetic axis from the original position (defined by the global minimum of the poloidal flux) over time for a 2D and 3D ITER simulation. $\tau_A$ is the Alfv\'en time.}
    \label{fig:axisshift}
\end{figure}

To test the agreement between the axisymmetric case and a fully 3D case, the same simulation was performed with resolution $128\times64\times32$.
While the equilibrium used in this test is prone to a $m=1$, $n=1$ magnetic kink instability, the growth time is $\sim 10^3 \tau_A$, which is significantly slower than the resistive wall time, and thus no dynamics in the toroidal direction are expected to take place before the VDE has matured (at which time the equilibrium and the associated kink instability are no longer applicable).
The shift in the position of the magnetic axis is shown in Figure \ref{fig:axisshift}, where strong agreement between the 2D and 3D cases is evident.

\section{Conclusions}
\label{sec:conclusions}

In this study, we have documented and verified the RW boundary BIM treatment implemented in the PIXIE3D MHD code for simulation of magnetically confined fusion devices. We have provided several analytical verification tests for the BIM module in realistic (but tractable) geometries that are accessible for any practitioner interested in testing their implementation without resorting to code-to-code comparisons (which are always fraught). Because of their analytical nature, our proposed RW BIM verification tests allow assessment of theoretical rates of convergence, a sensitive metric of correctness of implementation that is rarely possible otherwise. The appendices and supplemental data provide all relevant additional details needed for an end-to-end implementation of a BIM RW module in any other code. Our implementation has been demonstrated with a realistic VDE simulation in the ITER tokamak geometry, both in 2D and 3D, demonstrating its effectiveness.

\section*{Acknowledgments}
This work was supported by the U.S. Department of Energy Office of Fusion Energy Sciences and Office of Advanced Scientific Computing Research through the Tokamak Disruption Simulation (TDS) SciDAC project at Los Alamos National Laboratory (LANL) under Contract No. 89233218CNA000001. The research
used computing resources provided by the Los Alamos National Laboratory
Institutional Computing Program.
DB thanks Jim Bremer for his helpful communications regarding high order quadrature rules.

\appendix

\section{The boundary integral method for Laplace's equation}
\label{sec:BIM}
The boundary integral method (BIM) begins from Green's second identity,
\begin{align}
    \int_B \left(U\n\cdot\nabla W - W\n\cdot\nabla U \right) dS(\rr)
     = 
     \int\int_D\left(U\nabla^2 W - W\nabla^2 U\right) dV(\rr)
     \label{eq:greens_second}
\end{align}
where $B$ is a boundary enclosing some domain $D$, and $U$ is some scalar potential that satisfies Laplace's equation such that $\nabla^2U=0$ in $D$ and on $B$. In our case, $U$ is the magnetic scalar potential and $W$ is the Green's function $\G$, a fundamental solution to Laplace's equation satisfying
\begin{align}
    \nabla^2\G=\delta(\rr-\rp)
    \label{eq:greens-delta}
\end{align}
in $D$ and on $B$, where $\delta$ is the Dirac delta function. 
With (\ref{eq:greens-delta}) and $\nabla^2U=0$, the integral on the right hand side of (\ref{eq:greens_second}) is simply
\begin{align}
    \int\int_D U(\rr) \delta(\rr-\rp) dV(\rr) = U(\rp),
\end{align}
giving
\begin{align}
    \int_B \left(U(\rr)\n\cdot\nabla\G - \G\n\cdot\nabla U(\rr) \right) dS(\rr)
     = 
     U(\rp).
     \label{eq:greens-2}
\end{align}

The Green's function (\ref{eq:freespace-greens}) is singular when $\rr=\rp$ and therefore we must exclude this singular point from the domain of integration.
This is achieved by excluding the pole from the surface integral. This is described in \cite{hall1994a} and we briefly summarize here and give the result.
We are considering the external domain $D'$ of some enclosed volume $D$ with boundary $B$, with surface normal vector pointing towards the center of $D$, away from $D'$. The surface is broken into the hemispherical surface $C_\epsilon$ centered at the pole with radius $\epsilon$ and the majority surface $B-C_\epsilon$.
The sphere excluding the point $\rr=\rp$ has radius $\epsilon$ and differential surface area $dS=\epsilon^2d\Omega$, where $\Omega$ is the solid angle subtended by the $dS$. We are excluding the singular point from the surface by deforming the surface toward $D$, keeping the singular point within the domain of interest $D'$. Therefore, the radius of the sphere and \n~are aligned and we have
\begin{align*}
    \n\cdot\nabla = 
    \dfrac{\partial}{\partial\epsilon},
    \quad
    \epsilon(\rr) = |\rr-\rp|,
    \quad
    \G = -\dfrac{1}{4\pi\epsilon(\rr)},
\end{align*}
giving, for the integral over $C_\epsilon$ in the limit $\rr\rightarrow\rp$,
\begin{align*}
    &\lim_{\epsilon\rightarrow0}
    \int_{C_\epsilon}\left(
    \dfrac{U(\rr)}{4\pi\epsilon(\rr)^2}
    + 
    \dfrac{1}{4\pi\epsilon(\rr)}\dfrac{\partial U(\rr)}{\partial n}
    \right)
    \epsilon(\rr)^2d\Omega(\rr) \\
    &= 
    \lim_{\epsilon\rightarrow0}
    \int_{C_\epsilon}\left(
    \dfrac{U(\rr)}{4\pi}
    + 
    \dfrac{\epsilon(\rr)}{4\pi}\dfrac{\partial U(\rr)}{\partial n}
    \right)
    d\Omega(\rr) \\
    &= \dfrac{\Omega_e U(\rp)}{4\pi}.
\end{align*}
For a smoothly varying surface, the solid angle subtended by the hemisphere is $\Omega_e=2\pi$ and the resulting boundary integral equation (\ref{eq:greens-2}) is given in (\ref{eq:boundary_integral}). We note that the assumption of a smoothly varying surface imposes the restriction that any sharp features in the wall must be highly resolved.

\section{Green's Functions and their derivatives}
\label{sec:greens}
For the Green's function we work in cylindrical $(R,\phi,Z)$ coordinates, where the $R-Z$ plane describes the cross section of the torus and $\phi$ is the azimuthal angle (opposite to the toroidal azimuthal angle; $\phi_\mathrm{tor}=-\phi$).

In the cylindrical coordinate system, the distance between the two points $\rr$ and $\rp$ is
\begin{align*}
    |\rr-\rp|^2 &= R^2 + (R')^2 + (Z-Z')^2 -
    2RR'\cos(\phi-\phi') \\
    &= 2RR'(\chi-\cos(\phi-\phi')).
\end{align*}
The free space Green's function for Laplace's equation is therefore
\begin{align}
    \G &= -\dfrac{1}{4\pi\sqrt{2 RR'[\chi-\cos(\phi-\phi')]}},
    \label{eq:greens-rz-3d}
\end{align}
with derivatives
\begin{align*}
    \dfrac{\partial \G}{\partial R} &= 
    \dfrac{1}{8\pi R\sqrt{2RR'}}
    \left(\dfrac{1}{\sqrt{(\chi-\cos(\phi-\phi'))}}
    +\dfrac{R}{[\chi-\cos(\phi-\phi')]^\threehalf}
    \dfrac{\partial\chi}{\partial R}\right),\\
    \dfrac{\partial \G}{\partial Z}
    &= \dfrac{1}{8\pi\sqrt{2RR'}(\chi-\cos(\phi-\phi'))^\threehalf}
    \dfrac{\partial\chi}{\partial Z},
\end{align*}
and
\begin{align*}
    \dfrac{\partial\chi}{\partial R} &= 
    \dfrac{1}{R'} - \dfrac{R^2+(R')^2 + (Z-Z')^2}{2R^2R'}, \\
    \dfrac{\partial\chi}{\partial Z} &=
    \dfrac{Z - Z'}{RR'}.
\end{align*}

In this study, we will always assume the domain to be toroidally symmetric, although the solution can by either axisymmetric (2D) or asymmetric (3D). In the case that the solution is also assumed to be axisymmetric (2D), we can write:
\begin{align*}
    |\rr-\rp|^2 &= R^2 + (R')^2 - 2RR'\cos\phi + (Z-Z')^2 \\
    &=2RR'[\chi-\cos\phi],
\end{align*}
and therefore the free space Green's function for Laplace's equation (\ref{eq:freespace-greens}) is
\begin{align}
    \G &= -\dfrac{1}{4\pi\sqrt{2 RR'[\chi-\cos\phi]}}.
    \label{eq:greens-rz-2d}
\end{align}
With this imposed symmetry in the solution, (\ref{eq:bimRZ3d}) becomes
\begin{align}
% cylindrical
    \int U(R,Z)
    \left[
    (\n\cdot\er)
    \dfrac{\partial}{\partial R}
    \int_0^{2\pi}
    \G d\phi
    +
    (\n\cdot\ez)
    \dfrac{\partial}{\partial Z}
    \int_0^{2\pi}
    \G d\phi
    \right]
    R dR dZ  \nonumber\\
    - 
    \int
    \n\cdot\B(R,Z)
    \left(\int_0^{2\pi}
    \G ~ d\phi\right) R dR dZ
    &= \dfrac{U(R',Z')}{2}. \label{eq:bimRZ2d}
\end{align}
where we have used Leibniz's integral rule so that the integral of the Green's function over the rotational symmetry axis $\phi$ appears explicitly in both terms. This integral can be manipulated as
\begin{align}
    \Gint = \int_0^{2\pi}\G d\phi
    &= - \int_0^{2\pi}\dfrac{1}{4\pi\sqrt{2 RR'[\chi-\cos\phi]}} d\phi \nonumber\\
    &= \dfrac{4}{\sqrt{\chi+1}}\int_0^{\pi/2}
    (1-\mu^2\sin^2x)^{-\half}dx \nonumber\\
    &= -\dfrac{\mu K(\mu)}{2\pi\sqrt{RR'}}\label{eq:Gint}
\end{align}
We note that $\mu=\chi=1$ when $R=R'$ and $Z=Z'$. This occurs when the source and the target point in the Green's function are co-located. At this point, $K(\mu)$ diverges.
To complete the mathematical formulation of the axisymmetric boundary integral method, we require the following expressions for the derivatives
\begin{align}
    \dfrac{\partial \Gint}{\partial R} &=
    -\dfrac{\Gint}{2R} + \dfrac{\mu^3}{8\pi\sqrt{RR'}}
    \dfrac{E(\mu)}{(1-\mu^2)}
    \dfrac{\partial\chi}{\partial R},
    \nonumber\\
    \dfrac{\partial \Gint}{\partial Z} &= 
    \dfrac{\mu^3}{8\pi\sqrt{RR'}}
    \dfrac{E(\mu)}{(1-\mu^2)}
    \dfrac{\partial\chi}{\partial Z},
    \label{eq:Gderivs2d}
\end{align}
which completes the definition of the integrands in the boundary integral method.

\section{Elliptic integrals of the first, second and third kinds}
\label{sec:elliptic}
In this work, we encounter all three kinds of elliptic integrals. The incomplete elliptic integral of the first, second and third kinds are, respectively,
\begin{align}
    F(\Omega|\mu)&:=\int_0^\Omega
    \dfrac{d\phi}{\sqrt{\left(1-\mu^2\sin^2\phi\right)}},
    \quad\mu^2<1, \nonumber\\
    &= \int_0^{\sin\Omega}\dfrac{dx}{
    \sqrt{\left(1-x^2\right)}
    \sqrt{\left(1-\mu^2x^2\right)}
    }, \\
    E(\Omega,\mu)&:=\int_0^\Omega
    \sqrt{(1-\mu^2\sin^2\phi)}d\phi, \nonumber\\
    &= \int_0^{\sin\Omega}
    \dfrac{\sqrt{\left(1-\mu^2x^2\right)}}
    {\sqrt{\left(1-x^2\right)}} dx,
    \\
    \Pi(\Omega|n,m) &:= \int_0^\Omega 
    \dfrac{d\phi}{
    (1-n\sin^2\phi)
    \sqrt{1-m^2\sin^2\phi}
    },
    \nonumber\\
    &= 
    \int_0^{\sin\Omega}\dfrac{dx}{
    (1-nx^2)\sqrt{1-x^2}\sqrt{1-m^2x^2}}.
\end{align}
Their complete forms are obtained when $\Omega=\pi/2$:
\begin{align*}
    K(\mu) &= K(\pi/2|\mu) = \int_0^{\pi/2}
    (1-\mu^2\sin^2\phi)^{-\half}d\phi, \\
    E(\mu) &= E(\pi/2|\mu)= \int_0^{\pi/2}
    (1-\mu^2\sin^2\phi)^\half~d\phi. \\
    \Pi(n,m) &= \Pi(\pi/2|n,m).
\end{align*}
We also use the derivative of the complete elliptic integral of the first kind with respect to the parameter $\mu$, which is
\begin{align*}
    \dfrac{\partial K(\mu)}{\partial\mu} &=
    \dfrac{E(\mu)}{\mu(1-\mu^2)} - \dfrac{K(\mu)}{\mu}.
\end{align*}
We evaluate the elliptic integrals using the algorithm by \cite{fukushima2015a,fukushima2016a}.

\section{\label{sec:quadrature}Specialized Quadrature Rule for Singular and Hypersingular Integrands}
The integral of the Green's function over the interval containing the singularity ($\rr=\rp$) require special treatment because we were able to express the solution to the integral analytically. In other geometries and coordinate systems this is not always the case and integration of the singular and hypersingular integrands is made possible by constructing special Gaussian quadrature rules. Our approach is based on \cite{carley2007a}, but we also refer the reader to \cite{bremer2010a}. We use the Cauchy principal value integral and Hadamard finite part integral, defined by
\begin{align*}
    \mathcal{C}\int_a^b\dfrac{f(t)}{t-x}dt 
\end{align*}
and
\begin{align*}
    \mathcal{H}\int_a^b\dfrac{f(t)}{(t-x)^2}dt =
    -\dfrac{d}{dx}\left(\mathcal{C}\int_a^b\dfrac{f(t)}{t-x}dt\right),
\end{align*}
respectively.
Following \cite{carley2007a} and \cite{kolm2001a}, we sought to find the weights $w_j$ for each node $t_j$ of $N$-point quadrature rules such that the integral of the function $g(t)$ can be approximated as
\begin{align}
    \int_{-1}^1 g(t) dt &\approx \sum_{j=0}^{N-1} w_j g(t_j),
    \label{eq:lin_sys_quad}
\end{align}
where $g(t)$ is a linear combination of functions with singular terms up to $(x - t)^{-2}$.
In this work, we are mostly concerned with logarithmic and hypersingular integrands of type
\begin{align}
    g_1 = A(t)\ln|x-t| \label{eq:log_integrand_quadrature}\\
    g_2 = \dfrac{B(t)}{(x-t)^2} \label{eq:hyper_integrand_quadrature},
\end{align}
such that $g(t) = g_1 + g_2 + g_3$ where $A(t)$, $B(t)$ and $g_3$ are non-singular analytical functions in $t$ to which a truncated Taylor expansion of order up to $M-1$ converges in the range $[-1, 1]$. These will be represented using Legendre polynomials also of order up to $M-1$.
For a fixed value of the location of the singularity $x$, we solve the following linear system of equations for an $N$-point quadrature
\begin{align*}
    [\psi_{nj}] w_j = m_n,
    \quad j=0,\dots,N-1,\quad n = 0,\dots,4M-1
\end{align*}
where $m_n$ are the moments for the integrand,
\begin{align*}
    m_n &= \int_{-1}^1 P_n(t) dt \tag{$n=0,\ldots, M - 1$}\\
    m_n &= \int_{-1}^1 P_{n - M}(t)\ln|x-t| dt \tag{$n=M,\ldots, 2M - 1$} \\
    m_n &= \int_{-1}^1 \dfrac{P_{n - 2M}(t)}{(x-t)} dt\tag{$n=2M,\ldots, 3M - 1$} \\
    m_n &= \int_{-1}^1 \dfrac{P_{n - 3M}(t)}{(x-t)^2} dt,  \tag{$n=3M,\ldots, 4M - 1$}\\
\end{align*}
And $P_n(t)$ is the $n^\mathrm{th}$ order Legendre polynomial of the first kind. Although we are not explicitly dealing with terms with singularities of the form $(x - t)^{-1}$ we include them in our quadrature for completeness. The non-singular term has moments
\begin{align*}
    \int_{-1}^1P_ndt = 
    \left\{\begin{matrix}
    2,& \quad n=0\\
    0,& \quad n>1
    \end{matrix}
    \right. .
\end{align*}
For the logarithmic singularity, we evaluate the moments using
\begin{align*}
    \int_{-1}^1P_n\ln|x-t|dt = 
    \left\{\begin{matrix}
    (1-x)\ln(1-x) + (1+x)\ln(1+x) - 2,& \quad |x|\ne1, n=0\\
    \dfrac{2(Q_{n+1}(x) - Q_{n-1}(x))}{2n+1},& \quad
    |x|\ne 1, n>1
    \end{matrix}
    \right. .
\end{align*}
The singular integral the moments are evaluated by
\begin{align*}
    \int_{-1}^1\frac{P_n}{x - t}dt &= 
    P_n(x)\left(\mathcal{C}\int_{-1}^1\dfrac{1}{x-t}dt\right)
    +  \sum_{m=0}^{M-1}w_m\dfrac{P_n(t_m) - P_n(x)}{x-t_m},
\end{align*}
and those of the hypersingular integral by
\begin{align*}
    \int_{-1}^1\frac{P_n}{(x - t)^2}dt &= 
    P_n(x)\left(\mathcal{H}\int_{-1}^1\dfrac{1}{(x-t)^2}dt\right) \\
    &- 
    P_n'(x)\left(\mathcal{C}\int_{-1}^1\dfrac{1}{x-t}dt\right) \\
    &+  \sum_{m=0}^{M-1}w_m\dfrac{P_n(t_m) - P_n(x) +
    P_n'(x)(x-t_m)}{(x-t_m)^2}.
\end{align*}
We note that the equivalent formula for the hypersingular moments in \cite{carley2007a} has a typo in the quadrature term. $Q_n$ is the $n^\mathrm{th}$ order Legendre polynomial of the second kind, $w_m$ and $t_m$ are the weights and nodes of a standard $M$-point Gauss-Legendre quadrature. The first derivative of the Legendre polynomial $P$ is
\begin{align*}
    P_n'(x) = \left\{
    \begin{matrix}
    0, \quad &n=0, \\
    \dfrac{n}{x^2-1}(xP_n(x) - P_{n-1}(x)), \quad &n\ge1
    \end{matrix}
    \right.,
\end{align*}
and the Cauchy principal value and Hadamard finite part integrals are given by
\begin{align*}
    \mathcal{C}\int_{-1}^1\dfrac{1}{x-t}dt = 
    \ln\left|\dfrac{x+1}{1-x}\right|, \quad |x| \ne 1 \\
    \mathcal{H}\int_{-1}^1\dfrac{1}{(x-t)^2}dt = \dfrac{2}{x^2-1},
    \quad |x| \ne 1.
\end{align*}
We note that these formulae, and the quadrature rule we solve for, are valid for value of $x$ both within the domain of integration and outside of it (but not at $|x|=1$ without some additional formulae). This can prove useful when integrating over a domain that is close to, but does not contain, a singularity, when the integral behavior is \emph{near-singular}.
The matrix $\psi^k_{nj}$ is defined as
\begin{align*}
    \psi_{nj} &= P_n(t_j) \tag{$n=0,\ldots, M - 1$}\\
    \psi_{nj} &= P_{n - M}(t_j)\ln|x-t_j| \tag{$n=M,\ldots, 2M - 1$} \\
    \psi_{nj} &= \dfrac{P_{n - 2M} (t_j)}{x-t_j} \tag{$n=2M,\ldots, 3M - 1$}\\
    \psi_{nj} &= \dfrac{P_{n - 3M} (t_j)}{(x-t_j)^2} \tag{$n=3M,\ldots, 4M - 1$}.
\end{align*}
Provided that the functions $A(t)$ and $B(t)$ can be represented by an order $M-1$ polynomial, the quadrature rule should give the exact solution for sufficiently large $N$ ($N\ge4M$) up to near machine precision.
An important subtlety that may not be appreciated at first glance, is that even if the integrand can be represented exactly be a polynomial of degree 2, one still requires at least $4M$ nodes for the quadrature to be exact, where the weights were calculated for polynomials of up to order $M$. If the quadrature rule were computed with $M=8$, then at least 32 nodes would be needed to approximate the integral of $x^2$ to roundoff. In this work, for the singular integrands we found that $M=18$ was necessary, but for the regular integrands a standard 6-point Gauss-Legendre quadrature was sufficient, which greatly reduces computation time.

The system (\ref{eq:lin_sys_quad}) will be under- or over-determined if $N<M$ or $N>M$, respectively, and is well-posed when $N=M$. We found that the LAPACK minimum norm routine fails to solve the linear system correctly for underdetermined problems, but gives the correct solution for well-posed and for over-determined problems in the least-squares sense.
Quadrature nodes and weights for a 16-point rule are given in Table~\ref{tab:quad_nodes_weights}.

\begin{table}
    \caption{Nodes and weights for a 16-point quadrature rule for singular integrals of the form $A(t) + B(t)\ln|x-t| + C(t) / (x - t) + D(t) / (x - t)^2$ between -1 and 1, where $A(t)$, $B(t)$, $C(t)$ and $D(t)$ can be represented by a polynomial of order up to 3 and $x=0$.}
    \label{tab:quad_nodes_weights}
    \centering
    \begin{tabular}{cc}
        \mbox{Nodes} & \mbox{weights} \\
        \mbox{$x_j$} & \mbox{$w_j$} \\
        \hline
        -9.894009349916499E-01 & -1.474992869901327E+00\\
-9.445750230732326E-01 &  1.892638632193477E-01\\
-8.656312023878318E-01 &  1.767475890473514E+00\\
-7.554044083550030E-01 &  1.660052708947919E+00\\
-6.178762444026438E-01 & -6.839080653719604E-01\\
-4.580167776572274E-01 & -2.915680668468255E+00\\
-2.816035507792589E-01 &  2.664345793253724E+00\\
-9.501250983763745E-02 & -2.065566521529756E-01\\
 9.501250983763745E-02 & -2.065566521529759E-01\\
 2.816035507792589E-01 &  2.664345793253767E+00\\
 4.580167776572274E-01 & -2.915680668468290E+00\\
 6.178762444026438E-01 & -6.839080653719571E-01\\
 7.554044083550030E-01 &  1.660052708947910E+00\\
 8.656312023878318E-01 &  1.767475890473465E+00\\
 9.445750230732326E-01 &  1.892638632195361E-01\\
 9.894009349916499E-01 & -1.474992869901447E+00

     \end{tabular}
 \end{table}

\subsection{Verification of quadrature rule for integrals with logarithmic singularities}
We consider integrals of the form
\begin{align*}
    \int_{-1}^1t^n\ln|x-t|dt.
\end{align*}
For $n=0$,
\begin{align*}
    \int_{-1}^1t^n\ln|x-t|~dt &= \int_{-1}^1\ln|x-t|~dt\\
    &= \lim_{\epsilon=0}\left[
    \int_{\epsilon}^{1-x}\ln t dt + \int_\epsilon^{1+x}\ln t dt
    \right] \\
    &= (1-x)\ln(1-x) + (1+x)\ln(1+x) - 2.
\end{align*}
For $x=0$ and $n=0$,
\begin{align*}
    \int_{-1}^1t^n\ln|x-t|~dt = - 2,
\end{align*}
the numerical error for which is at roundoff (Table \ref{tab:quadresults_ln1}).
\begin{table}
     \caption{Relative error for quadrature of $\ln|-t|$.}
     \label{tab:quadresults_ln1}
     \centering
    \begin{tabular}{cc}
        \mbox{Nodes} & \mbox{relative error} \\
        \hline
        16  & 2.2204460492503126E-016 \\
%        20  & 2.2204460492503126E-016 \\
%        24  & 2.2204460492503126E-016 \\
%        28  & 4.4408920985006281E-016 \\
        32  & 2.2204460492503126E-016 \\
%        36  & 0.0000000000000000 \\
%        40  & 6.6613381477509353E-016 \\
%        44  & 5.5511151231257857E-016 \\
%        48  & 2.2204460492503126E-016 \\
%        52  & 0.0000000000000000 \\
%        56  & 4.4408920985006242E-016 \\
%        60  & 2.2204460492503126E-016 \\
        64  & 8.8817841970012444E-016
     \end{tabular}
 \end{table}
For $x=0$ and $n=2$,
\begin{align*}
    \int_{-1}^1 t^n\ln|x-t| dt &= \int_{-1}^1 t^2\ln|-t| dt \\
    &= 2\int_0^1 t^2\ln t dt \\
    &= 2 \lim_{\epsilon=0}\int_\epsilon^1 t^2\ln t dt \\
    &= -\dfrac{2}{9},
\end{align*}
which we show in Table~\ref{tab:quadresults_hyper2} is obtained to roundoff using the quadrature.
\begin{table}
     \caption{Relative error for quadrature of $t^2\ln|-t|$.}
     \label{tab:quadresults_hyper2}
     \centering
    %\begin{ruledtabular}
    \begin{tabular}{cc}
        Nodes & relative error \\
        \hline
   16 &  1.1241008124329698E-015 \\
%   20 &  1.2490009027033011E-016 \\
%   24 &  8.7430063189231009E-016 \\
%   28 &  6.2450045135165095E-016 \\
   32 &  7.4940054162198017E-016 \\
%   36 &  1.6237011735142889E-015 \\
%   40 &  1.2490009027032995E-015 \\
%   44 &  1.1241008124329724E-015 \\
%   48 &  1.2490009027033014E-016 \\
%   52 &  2.4980018054066017E-016 \\
%   56 &  4.9960036108132074E-016 \\
%   60 &  1.2490009027033027E-015 \\
   64 &  1.8735013540549481E-015
     \end{tabular}
    %\end{ruledtabular}
\end{table}

\subsection{Verification of quadrature rule for hypersingular integrals}
For $n=0$:
\begin{align*}
    \int_{-1}^1\dfrac{1}{(x-t)^2}dt = -\dfrac{1}{x-1} + \dfrac{1}{x+1}
\end{align*}
and when $x=0$ and $n=0$:
\begin{align*}
    \int_{-1}^1\dfrac{1}{(x-t)^2}dt = 2.
\end{align*}
For $n=2$:
\begin{align*}
    \mathcal{H}\int_{-1}^1\dfrac{t^2}{(x-t)^2}dt &=
    -\dfrac{d}{dx}\left(\mathcal{C}\int_{-1}^1\dfrac{t^2}{x-t}dt\right)\\
    &=-\dfrac{d}{dx}\left(\lim_{\epsilon=0}\left[
    \int_{-1}^{x-\epsilon}\dfrac{t^2}{x-t}dt + 
    \int_{x+\epsilon}^1\dfrac{t^2}{x-t}dt
    \right]\right) \\
    &= \dfrac{4x^2 - 2(x^2-1)x\ln(-\frac{x+1}{x-1})-2}
    {x^2-1}
\end{align*}
trivially giving, when $x=0$ and $n=2$:
\begin{align*}
    \mathcal{H}\int_{-1}^1\dfrac{t^2}{(0-t)^2}dt = 2
\end{align*}
and when $x=\frac{1}{2}$ and $n=2$:
\begin{align*}
    \mathcal{H}\int_{-1}^1\dfrac{t^2}{(\frac{1}{2}-t)^2}dt = 
    \frac{4}{3} - \ln3
\end{align*}

\begin{table}
     \caption{Relative error for quadrature of $t^2/(\frac{1}{2}-t)^2$.}
     \label{tab:my_label}
     \centering
    %\begin{ruledtabular}
    \begin{tabular}{cl}
        \mbox{Nodes} & \mbox{relative error} \\
        \hline
%         4  &  5.6759615715343042E-015\\
%         8  &  4.2285913707912444E-013\\
%         12 &   1.8919871905114275E-015\\
         16 &   1.0760677146033840E-014\\
%         20 &   2.5920224510013226E-013\\
%         24 &   6.5037059673830621E-015\\
%         28 &   2.4359335077833990E-014\\
         32 &   8.3554884300890943E-013\\
%         36 &   6.3145072483314786E-014\\
%         40 &   8.1473698391391554E-014\\
%         44 &   9.8466108346275881E-013\\
%         48 &   5.5813622120083893E-014\\
%         52 &   2.1970701249818738E-013\\
%         56 &   5.9562121741252290E-013\\
%         60 &   8.6203666367684178E-014\\
         64 &   2.5092480114164056E-013
     \end{tabular}
     %\end{ruledtabular}
\end{table}

\subsection{Elliptic Integrals}
In the axially symmetric representation of the toroidal resistive wall BIM problem, we encounter complete elliptic integrals of the first and second kind. The complete elliptic integral of the second kind, $E(\mu)$, is regular and its term appears as
\begin{align*}
    \int_{1-\delta}^1\dfrac{E(\mu)}{(1-\mu^2)}d\mu =
    \int_{1-\delta}^1\dfrac{E(\mu)}{(1+\mu)(1-\mu)}d\mu,
\end{align*}
and therefore is taken care of by the singular term $P_n(t)/(x-t)$ in the specialized quadrature rule in this Appendix.

The complete elliptic integral of the first kind, $K(\mu)$, is divergent at $\mu=1$ but $(1-\mu)K(\mu)$ is regular over the interval $\mu = {0,1}$. Therefore, it is also taken care of by the singular term $P_n(t)/(x-t)$ in the specialized quadrature rule in this Appendix. Indeed,
\begin{align*}
    \int_{1-\delta}^1 K(\mu) d\mu = 
    2 - 2E(1-\delta) + 2\delta K(1-\delta).
\end{align*}

Complete elliptic integrals of the third kind also exhibit divergent singular behavior and appear to be more problematic at first glance, however they, too, become regular when muliplied with $(1-\mu)$.

\section{Biot-Savart law}
\label{sec:biot-savart}
For an axisymmetric filament source in cylindrical coordinates,
\begin{align*}
d\lb\times\rp &= (Z_r-Z_l)R_ld\phi_l\cos\phi_l\ex + (Z_r-Z_l)R_ld\phi_l\sin\phi_l\ey
+ R_ld\phi_l(R_l-R_r\cos(\phi_l-\phi_r))\ez\\
&= (Z_r-Z_l)d\phi_lR_l\cos(\phi_r-\phi_l)\er
- (Z_r-Z_l)d\phi_lR_l\sin(\phi_r-\phi_l)\ephi \\
&+ R_ld\phi_l(R_l-R_r\cos(\phi_l-\phi_r))\ez
\end{align*}
where subscripts $l$ and $r$ indiate to the source and target, respectively.
Therefore, for an axially symmetric source coil and target surface the magnetic field is
\begin{align*}
    \B(\rr)\cdot\er &= \dfrac{\mu_0I}{4\pi(2RR_l)^\threehalf}(Z-Z_l)R_l
    \int_0^{2\pi}\dfrac{\cos(\phi-\phi_l)d\phi_l}{[\chi-\cos(\phi-\phi_l)]^\threehalf},\\
    \B(\rr)\cdot\ephi &= \dfrac{\mu_0I}{4\pi(2RR_l)^\threehalf}(Z-Z_l)R_l
    \int_0^{2\pi}\dfrac{\sin(\phi-\phi_l)d\phi_l}{[\chi-\cos(\phi-\phi_l)]^\threehalf}=0,\\
    \B(\rr)\cdot\ez &= \dfrac{\mu_0I}{4\pi(2RR_l)^\threehalf}\left[
    R_l^2\int_0^{2\pi}\dfrac{d\phi_l}{[\chi-\cos(\phi-\phi_l)]^\threehalf}
    -R_rR_l\int_0^{2\pi}\dfrac{\cos(\phi_l-\phi)d\phi_l}{[\chi-\cos(\phi-\phi_l)]^\threehalf}
    \right].
\end{align*}
The integrals can be expressed using the complete elliptic integrals of the first, second and third kinds as follows.
\begin{align}
    \label{eq:biot-savart-toroidal-br}
    \B(\rr)\cdot\er &= \dfrac{\mu_0I}{4\pi(2RR_l)^\threehalf}
    \dfrac{2(Z-Z_l)R_l}{(\chi+1)^\threehalf}
    \int_\frac{\pi}{2}^\frac{3\pi}{2}
    \dfrac{\cos(2\varphi-\pi)d\varphi}{(1-\mu^2\sin^2\varphi)^\threehalf} ,\nonumber\\
    &= \dfrac{\mu_0I}{\pi(2RR_l)^\threehalf}
    \dfrac{(Z-Z_l)R_l}{\mu^2(\chi+1)^\threehalf}\left[
    2K(\mu) - \dfrac{(\mu^2-2)E(\mu)}{\mu^2-1}
    \right],\\
    \B(\rr)\cdot\ez &= -\dfrac{\mu_0I}{4\pi(2RR_l)^\threehalf}
    \dfrac{4R_l}{(\chi+1)^\threehalf}
    \left[
    R_l\int_0^\frac{\pi}{2}
    \dfrac{d\varphi}{(1-\mu^2\sin^2\varphi)^\threehalf}
    +R\int_0^\frac{\pi}{2}
    \dfrac{\cos(2\varphi)d\varphi}{(1-\mu^2\sin^2\varphi)^\threehalf}
    \right]\nonumber\\
    &=-\dfrac{\mu_0I}{\pi(2RR_l)^\threehalf}
    \dfrac{R_l}{(\chi+1)^\threehalf}
    \left[
    (R_l+R)\Pi(\mu,\mu^2) + \dfrac{2R}{\mu^2}\left(
    K(\mu) + \dfrac{E(\mu)}{\mu^2-1}\right)
    \right].
    \label{eq:biot-savart-toroidal-bz}
\end{align}

\bibliographystyle{elsarticle-num}
\bibliography{refs}

\end{document}